\documentclass[12pt,a4paper]{article}
\usepackage{a4wide}

\usepackage[english]{babel}
\usepackage[utf8]{inputenc}

\usepackage{graphicx}
\usepackage{axodraw}
\usepackage[numbers,sort&compress]{natbib}
\usepackage{amsfonts,amsmath,amssymb}
\usepackage[ps2pdf,bookmarks,colorlinks,bookmarksnumbered]{hyperref}

\newcommand{\vect}[2][c]{\left( \hspace{-3pt}
\begin{array}{#1}
	#2
\end{array}
\hspace{-2pt} \right)}

\newcommand{\phys}{\mathrm{phys}}

\newcommand{\GeV}{\text{GeV}}

\renewcommand{\L}{\mathcal{L}}
\renewcommand{\O}{\mathcal{O}}
\newcommand{\A}{\mathcal{A}}
\newcommand{\M}{\mathcal{M}}

\begin{document}

\begin{titlepage}
\begin{flushright}
\small
LU TP 12-02\\
revised March 2012
\end{flushright}

\vfill

\begin{center}
{\LARGE\bf Leading logarithms in the anomalous\\sector of two-flavour QCD}
\\[2cm]
{\bf Johan Bijnens$^a$, Karol Kampf$\hskip0.2ex^{b,a}$ and Stefan
Lanz$^{a}$}\\[2mm]
{$^a\,$Department of Astronomy and Theoretical Physics, Lund University}\\
{S\"olvegatan 14A, S 223 62 Lund, Sweden}\\[2mm]
{$^b\,$Institute of Particle and Nuclear Physics,\\Faculty of
Mathematics and Physics,}\\
{Charles University, 18000 Prague, Czech Republic.}
\end{center}

\vfill

\begin{abstract}
We add the Wess-Zumino-Witten term to the $N=3$ massive nonlinear sigma model
and study the leading logarithms in the anomalous sector.
We obtain the leading logarithms to six loops for $\pi^0\to\gamma^*\gamma^*$
and to five loops for $\gamma^*\pi\pi\pi$. In addition we extend the earlier work
on the mass and decay constant to six loops and the vector form factor to
five loops. We present numerical result for the anomalous processes and the
vector form factor.  In all cases the series are found to converge rapidly.

\vspace{3ex}

\end{abstract}

\vfill

\end{titlepage}

\tableofcontents

\section{Introduction}

Obtaining exact results in quantum field theory is rather difficult.
One of the few things which can be easily calculated to all orders in
renormalizable theories are the leading logarithms of the type
$(g^2\log\mu^2)^n$ where $\mu$ is the subtraction scale and $g$ the coupling
constant. The analogue of this in effective field theories is not so simple
since at each order in the expansion new terms in the Lagrangian appear
and the recursive argument embedded in the renormalization group equations
for renormalizable theories no longer applies. Nonetheless, one can
calculate the leading logarithms in effective field theories using only
one-loop calculations. This was suggested at two-loop order by Weinberg
\cite{Weinberg:1978kz} and proven to all orders in \cite{Buchler:2003vw}.

In the massless case this has been used to very high orders in
\cite{Kivel:2008mf,Kivel:2009az,Koschinski:2010mr} for meson-meson scattering and
form factors. In the massive case, many more terms contribute but
in \cite{Bijnens:2009zi,Bijnens:2010xg} a method was developed for
handling those, and the leading logarithms in the massive nonlinear sigma model
were obtained to five-loop order for the mass, decay constant and the
vacuum expectation value and to four-loop order for the vector and scalar
form factors and the meson-meson scattering amplitude.
A natural continuation of that program is to extend it to other sectors
as well. We therefore add to the massive nonlinear sigma model
for $N=3$ the anomalous part via the Wess-Zumino-Witten (WZW) term. This allows
us to study the leading logarithms for anomalous processes in two-flavour
Chiral Perturbation Theory (ChPT). In addition one can hope that in this sector with
its many nonrenormalization theorems it might be easier to guess the all order
results when the first terms in the series are known.
The WZW term only makes sense for $N=3$ so we do not work out the results
for general $N$ in this case.

We have also improved the programs used in
\cite{Bijnens:2009zi,Bijnens:2010xg} so that they now can be used to
arbitrarily high orders given enough computing power. So, at least in principle,
the problem of the leading logarithms is solved.
In practice we obtained one order more than in the earlier work.

The main part of the paper is devoted to the calculations of the
leading logarithms (LL) for the two main anomalous processes in the pion
sector, the full $\pi^0\gamma^*\gamma^*$ and $\gamma^*\pi\pi\pi$ vertices.
For the former we obtained the LL to six loops and for the latter to five.
The results indicate that in all cases the chiral expansion converges fast but
we did not find a simple all-order conjecture.

The results agree with all known relevant earlier calculations. As
an additional check we have used several different parametrizations of the
fields.

In Sect.~\ref{sigmamodel} we introduce shortly the massive nonlinear sigma model
and in Sect.~\ref{sectWZW} the two-flavour Wess-Zumino-Witten term.
Sect.~\ref{sec:LLeven} contains the discussion of LL in the nonanomalous sector
where we present our new results and show some numerical results.
Here we also explain briefly the principles of the calculation.
More details on the method can be found
in \cite{Bijnens:2009zi,Bijnens:2010xg}.
Sects.~\ref{secLLodd} and \ref{secpheno} are the main part of this paper.
The LL are calculated in Sect.~\ref{secLLodd}. 
We did not find a simple all-order conjecture but the LL indicate for example
that the nonfactorizable part in $\pi^0\gamma^*\gamma^*$ with both
photons off shell should be small.
We present numerical results in Sect.~\ref{secpheno}. In all cases we
find good convergence. Sect.~\ref{conclusions} shortly recapitulates our
results. The appendix contains a dispersive argument to clarify the discrepancy
with \cite{Kivel:2009az} for the vector form factor.

\section{The model}

\subsection{Massive nonlinear 
 \texorpdfstring{$O(N+1)/O(N)$}{O(N+1)/O(N)} sigma model}
\label{sigmamodel}

The $O(N+1)/O(N)$ nonlinear sigma model, including external sources,
is given by the Lagrangian
\begin{equation}
\label{sigmalag}
\L_{n\sigma} = \frac{F^2}{2}
D_\mu \Phi^T D^\mu \Phi+F^2 \chi^T \Phi\,.
\end{equation}
$\Phi$ is a real $N+1$ vector,
$\Phi^T =\left(\Phi^0~\Phi^1~\ldots~\Phi^N\right)$,
which  satisfies the constraint $\Phi^T\Phi=1$
and transforms under the fundamental
representation of $O(N+1)$.
The covariant derivative is
\begin{eqnarray}
D_\mu \Phi^0 &=& \partial_\mu \Phi^0 + a^a_\mu \Phi^a\,,
\nonumber\\
D_\mu \Phi^a &=& \partial_\mu\Phi^a +v^{ab}_\mu\Phi^b-a^a_\mu\Phi^0\,.
\end{eqnarray}
The vector sources are antisymmetric, $v^{ab}_\mu=-v^{ba}_\mu$, and correspond
to the unbroken group generators. The axial sources $a^a_\mu$ correspond
to the broken generators.
Lower-case Latin indices $a,b,\ldots$ run over $1,\ldots,N$ in the remainder
and are referred to as flavour indices.
The mass term
$\chi^T \Phi$
contains the scalar, $s^0$, and pseudoscalar, $p^a$, external sources as well
as the explicit symmetry breaking term $M^2$:
\begin{equation}
\chi^T =\left(~(2Bs^0+M^2)~~p^1~\ldots~p^N\right)\,.
\end{equation}
The vacuum condensate
\begin{equation}
\label{vacuum}
\langle \Phi^T \rangle = \left(1~0~\ldots~0\right)\,
\end{equation}
breaks $O(N+1)$ spontaneously to $O(N)$.
We thus have in principle $N$ Goldstone bosons represented by $\phi$.
The explicit symmetry breaking term, the part containing $M^2$,
breaks the $O(N+1)$ symmetry to $O(N)$, enforcing 
the vacuum condensate to be in the direction~(\ref{vacuum}) and gives
a mass to the Goldstone bosons which at tree level is exactly $M$.

This particular model is the same as lowest-order two-flavour
ChPT for $N=3$~\cite{Gasser:1983yg,Weinberg:1968de}
and has been used as a model for strongly interacting Higgs sectors
in several scenarios; see, e.g.,~\cite{Einhorn:1984mr,Sannino:2008ha}.

The terminology for the external sources or fields is taken from
two-flavour ChPT. The vector currents for $N=3$ are given by
$v^{ab} = -\varepsilon^{abc} v^c$ with $\varepsilon^{abc}$
the Levi-Civita tensor. The electromagnetic current
at lowest order is associated with $v^3$.

We write $\Phi$ in terms of a real $N$-component vector $\phi$,
which transforms linearly under
the unbroken part of the symmetry group $O(N)$.
We have made use of five different parametrizations in order
to check the validity of our results. They are
\begin{eqnarray}
\Phi_1 &=&
\vect{\sqrt{1-\frac{\phi^T\phi}{F^2}}\\ \frac{\phi}{F}}\,,
\hskip2cm
\Phi_2 = \frac{1}{\sqrt{1+\frac{\phi^T\phi}{F^2}}}
\vect{1\\\frac{\phi}{F}}\,,
\nonumber\\[5mm]
\Phi_3 &=& \vect{1-\frac{1}{2}\frac{\phi^T\phi}{F^2}\\[2mm]
\sqrt{1-\frac{1}{4}\frac{\phi^T\phi}{F^2}}\,\frac{\phi}{F}}\, ,
\hskip1.2cm
\Phi_4 = \vect{\cos\sqrt{\frac{\phi^T\phi}{F^2}}
\nonumber\\[2mm]
\sin\sqrt{\frac{\phi^T\phi}{F^2}}\,\frac{\phi}{\sqrt{\phi^T\phi}}}
\, , 
\nonumber\\[5mm]
\Phi_5 &=& \frac{1}{1+\frac{\phi^T\phi}{4 F^2}}
\vect{1-\frac{\phi^T\phi}{4 F^2}\\[2mm] \frac{\phi}{F}} \, .
\label{parametrisations}
\end{eqnarray}
$\Phi_1$ is the parametrization used in~\cite{Gasser:1983yg},
$\Phi_2$ a simple variation.
$\Phi_3$ is such that the explicit symmetry breaking term in~(\ref{sigmalag})
only gives a mass term to the $\phi$ field but no vertices.
$\Phi_4$ is the parametrization one ends up with if using the general
prescription of~\cite{Coleman:1969sm}. Finally, $\Phi_5$ has been used by
Weinberg in, e.g.,~\cite{Weinberg:1968de,Weinberg:1978kz}. The reason for
using different parametrizations is that for each one of them, the contributions
are distributed very differently between the diagrams and obtaining the same
result thus provides a thorough check on our calculations.

\subsection{Wess-Zumino-Witten Lagrangian}
\label{sectWZW}

For $N=3$, the massive nonlinear $O(N+1)/O(N)$ sigma model corresponds to
two-flavour chiral perturbation theory, which is an effective field theory
for QCD. It is well known that the
chiral axial current of the latter is
anomalous~\cite{Adler:1969gk,Adler:1969er,Bardeen:1969md,Bell:1969ts},
leading to the occurrence of processes such as
$\pi^0 \to \gamma \gamma$ or $\pi \gamma \to \pi \pi$. The Lagrangian that
we have introduced above does, however, not account for this.
The necessary interaction terms are contained in the Wess-Zumino-Witten
term~\cite{Wess:1971yu,Witten:1983tw},
which must be added to the effective Lagrangian. It is constructed such that
it reproduces the anomalous Ward identities.
Kaiser derived the WZW term for two-flavour ChPT~\cite{Kaiser:2000ck}, where it
is considerably simpler than in the case of three flavours. His result can
be re-expressed in terms of the field $\Phi$ as
\begin{eqnarray}
\L_\textit{WZW} &=&-\frac{N_c}{8 \pi^2} \epsilon^{\mu \nu \rho \sigma}
  \bigg\{
\epsilon^{abc} \bigg(
   \frac{1}{3} \Phi^0 \partial_\mu \Phi^a \partial_\nu \Phi^b \partial_\rho \Phi^c
- \partial_\mu \Phi^0 \partial_\nu \Phi^a \partial_\rho \Phi^b \Phi^c
          \bigg) v_\sigma^0 \nonumber \\[1mm]
&&+ ( \partial_\mu \Phi^0 \Phi^a - \Phi^0 \partial_\mu \Phi^a )
         v_\nu^a \partial_\rho v_\sigma^0
+ \frac{1}{2} \epsilon^{abc} 
    \Phi^0 \Phi^a  v_\mu^b v_\nu^c \partial_\rho v_\sigma^0
\bigg\} \, .
\label{LWZW}
\end{eqnarray}
The interaction with the axial current coming from the WZW term
has been omitted. Note that the normalization of the vector field differs 
from the one used in~\cite{Kaiser:2000ck}.
The Lagrangian depends on the Levi-Civita tensor $\epsilon^{abc}$
in the $SO(3)$ flavour indices. This is an object specific to $N=3$. There
is no obvious simple generalization\footnote{The generalization to
different $n$ for the $SU(n)\times SU(n)/SU(n)$ case is easy but that is a 
different case than the $O(N+1)/O(N)$ considered here.}
to different $N$ so for anomalous
processes we restrict the calculation to $N=3$.

The Lagrangian in~(\ref{LWZW}) is of chiral order $p^4$ implying
that anomalous processes are of the same order at leading order,
while one-loop corrections are already $\O(p^6)$.
This is immediately clear from the presence of the epsilon tensor with
four Lorentz indices:
each one of them must be combined with either a derivative or an
external vector field, both of which are $\O(p)$.

The interaction of the pseudoscalars with the photon field $\A_\mu$ can be
obtained from the WZW Lagrangian by
setting the vector current to
\begin{equation}
 v_\mu^0 = \frac{e}{3} \A_\mu \, , \qquad v_\mu^a = e \A_\mu \delta^{a3} \, .
\end{equation}

The Lagrangian of~(\ref{sigmalag}) has also a symmetry that QCD does not
have~\cite{Witten:1983tw}. The fields under this extra symmetry,
called intrinsic parity, transform as
\begin{equation}
\Phi^0\to \Phi^0,\quad \Phi^a\to -\Phi^a,\quad v^{ab}_\mu \to v^{ab}_\mu,
\quad a^a_\mu\to-a^a_\mu,\quad s\to s,\quad p\to -p\,.
\end{equation}
The Lagrangian~(\ref{sigmalag}) and higher orders are even under this symmetry
while~(\ref{LWZW}) is odd.

As in the even sector, we have used several of the parametrizations
given in~(\ref{parametrisations}). Intrinsic parity for the different $\phi$
is such that it is always odd, $\phi^a\to-\phi^a$.
The WZW term leads to interactions of one, two, or three photons with
an odd number of pions. The anomaly also generates purely mesonic interactions
among an odd number of five or more Goldstone bosons. However, for two flavours
the purely mesonic odd intrinsic parity processes vanish to all orders.

We use anomalous and odd intrinsic parity as synonyms and refer to the
$N=3$ case here occasionally as the two-flavour case since it corresponds
to two light quark flavours.

\section{Leading logarithms in the even sector}
\label{sec:LLeven}

In this section we recapitulate and extend some results of the
even intrinsic parity sector of~\cite{Bijnens:2009zi,Bijnens:2010xg}.
We focus on those results that will be needed for the later calculations
in the anomalous sector.

The leading dependence on $\log\mu$ at each order, with $\mu$ the subtraction scale,
is what we call the leading logarithm.
It can in principle always be obtained from one-loop calculations
as was proven using $\beta$-functions in~\cite{Buchler:2003vw}
and in a simpler diagrammatic way in~\cite{Bijnens:2009zi}.
We will discuss some of those results in the sections on the explicit calculation of the
mass and $\gamma\pi\to\pi\pi$.

In effective field theories there is a new Lagrangian at every order.
The observation of~\cite{Bijnens:2009zi} that the needed parts of
those Lagrangians can be generated automatically from the
one-loop diagrams allowed to perform the calculations.
The actual calculations were performed by
using FORM~\cite{Vermaseren:2000nd} extensively.

We have extended the programs used in~\cite{Bijnens:2009zi,Bijnens:2010xg}
so that they can in principle run to an arbitrary order\footnote{In~\cite{Bijnens:2009zi,Bijnens:2010xg},
a different program was used for each diagram with a given number of vertices/propagators
and the list of possible diagrams was constructed by hand.}.
The only limit is set by computing
time, which grows rapidly with the order.
The necessary additions include routines that
generate the required diagrams at a given order
and calculate one-loop diagrams with an arbitrary number of propagators.
In this way, we have verified some of the earlier results
and obtained the coefficient of one more order for the mass, decay constant
and vector form factor.

In effective field theories writing the expansion in terms of lowest
order or physical quantities can make quite a big difference
in the rate of convergence. We therefore follow~\cite{Bijnens:2010xg}
in using two different expansions
in terms of leading logarithms.
A given observable $O_\phys$ can be written in different ways:
\begin{eqnarray}
\label{defexpL}
O_\phys &=&  O_0 \left(1+a_1 L +a_2 L^2+\cdots\right)\,,
\\
\label{defexpLphys}
O_\phys &=&  O_0 \left(1+c_1 L_\phys +c_2 L_\phys^2+\cdots\right)\,,
\end{eqnarray}
where the chiral logarithms are defined either from the lowest-order parameters
$M$ and $F$ as
\begin{equation}
\label{defL}
 L \equiv \frac{M^2}{16\pi^2F^2} \log \frac{\mu^2}{M^2} \,,
\end{equation}
or from the physical decay constant $F_\pi$ and mass $M_\pi$ as
\begin{equation}
\label{defLphys}
 L_\phys \equiv \frac{M_\pi^2}{16\pi^2F_\pi^2} \log \frac{\mu^2}{M_\pi^2}\,.
\end{equation}

\subsection{Mass}
\label{secmass}

In this subsection we will present the first few coefficients $a_i$ and $c_i$
of the expansions in~(\ref{defexpL}) and~(\ref{defexpLphys}) with
$O_\phys = M_\pi^2$ and $O_0 = M^2$. In addition, 
we have also calculated the generic two-point function, which
is needed for the wave-function renormalization.
We extend the result from~\cite{Bijnens:2009zi} by giving the expansion
coefficients $a_i$ and $c_i$ also at sixth order.

Let us briefly recapitulate the strategy that is followed to obtain
the expansion coefficients.
The starting point is the Lagrangian~(\ref{sigmalag}) which generates
vertices with an arbitrary even
number of pion legs. These lowest-order vertices are diagrammatically
denoted by
\begin{picture}(11,0)(0,0)
\SetScale{1.0}
\SetPFont{Helvetica}{7}
\BText(5,4){0}
\end{picture}
with the corresponding number of legs appended. The zero in this symbol
refers to the order in the expansion. If we want to calculate the leading
logarithm for the two-point function at one-loop level,
we must evaluate the tadpole diagram with one insertion of the leading-order
four-pion vertex. Its divergence must be canceled by a counter term from the next order.
This can be depicted schematically as
\begin{equation}
\label{massll1}
\parbox{50mm}{
\begin{picture}(40,60)(0,0)
\SetScale{1.0}
\SetPFont{Helvetica}{7}
\BCirc(40,40){20}
\Line(40,20)(49.1645,2.22329)
\Line(40,20)(30.8355,2.22329)
\BText(40,20){0}
\Text(80,18)[]{$\Longrightarrow$}
\Line(110,20)(150,20)
\BText(130,20){1}
\end{picture}}
\end{equation}
Also at the next order, divergences must cancel,
but the situation is somewhat more complicated. There are two one-loop diagrams
from which, using the results of~\cite{Bijnens:2009zi,Buchler:2003vw},
the leading divergence can be determined. This thus determines the
relevant part of the second-order Lagrangian:
\begin{equation}\label{massll2}
\parbox{100mm}{\begin{picture}(80,65)(0,0)
\SetScale{1.0}
\SetPFont{Helvetica}{7}
\BCirc(40,40){20}
\Line(40,20)(49.1645,2.22329)
\Line(40,20)(30.8355,2.22329)
\BText(40,20){1}
\put(80,18){,}
\end{picture}
\begin{picture}(80,60)(0,0)
\SetScale{1.0}
\SetPFont{Helvetica}{7}
\BCirc(40,40){20}
\Line(40,20)(49.1645,2.22329)
\Line(40,20)(30.8355,2.22329)
\BText(40,20){0}
\BText(40,60){1}
\Text(80,18)[]{$\Longrightarrow$}
\Line(110,20)(150,20)
\BText(130,20){2}
\end{picture}}
\end{equation}
We already have all information ready in order to calculate the second diagram:
the tree-level vertex follows directly from the lowest-order Lagrangian
and the next-to-leading-order
vertex has been determined in~(\ref{massll1}).
The first diagram, however, contains
the next-to-leading-order vertex with four pion legs, which 
we do not know yet. In order
to obtain its divergence, we must calculate two more diagrams:
\begin{equation}
\parbox{100mm}{
\begin{picture}(80,75)(0,0)
\SetScale{1.0}
\SetPFont{Helvetica}{7}
\BCirc(40,40){20}
\Line(40,20)(55.115,6.90279)
\Line(40,20)(45.6347,0.810141)
\Line(40,20)(34.3653,0.810141)
\Line(40,20)(24.885,6.90279)
\BText(40,20){0}
\put(80,18){,}
\end{picture}
\begin{picture}(80,60)(0,0)
\SetScale{1.0}
\SetPFont{Helvetica}{7}
\BCirc(40,40){20}
\Line(40,20)(49.1645,2.22329)
\Line(40,20)(30.8355,2.22329)
\BText(40,20){0}
\Line(40,60)(30.8355,77.7767)
\Line(40,60)(49.1645,77.7767)
\BText(40,60){0}
\Text(80,18)[]{$\Longrightarrow$}
\Line(110,2)(150,38)
\Line(110,38)(150,2)
\BText(130,20){1}
\end{picture}}
\end{equation}
The algorithm continues to higher orders in exactly the same way.
All the diagrams
needed for the two-point function up to third order are shown in
\cite{Bijnens:2009zi}, Figs.~3--5. The total number of diagrams needed
for the mass up to order $n$ is $1,5,16,45,116,303,\ldots$.

We did not find a simple formula to estimate the total number of diagrams
needed. We did find a conjecture, verified up to 12th order, about the number
of diagrams needed with only two external legs at each order:
\begin{equation}
 \text{\# two-point diagrams} = \begin{cases}
2^{n-2} + 3\times 2^{\tfrac{n-3}{2}} - 1 & \mbox{for } n\mbox{ odd} \\
2^{n-2} + 2^{\tfrac{n}{2}} -1, &\mbox{for } n\mbox{ even}
\end{cases} \, ,
\end{equation}
that is: 1, 2, 4, 7, 13, 23, 43, 79, 151, \ldots\, diagrams with two external
legs at order $n$.

\begin{figure}[t]
\begin{minipage}{0.49\textwidth}
\includegraphics[width=\textwidth,clip=true,trim=1.6cm 0 1.2cm 0]{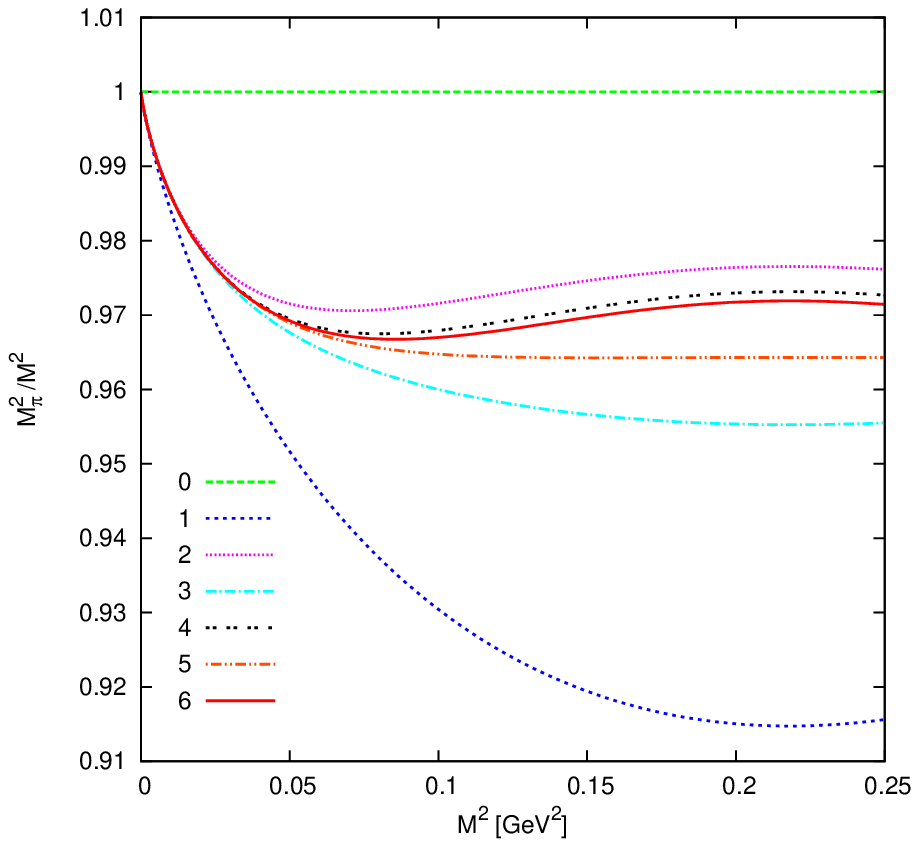}
\end{minipage}
\begin{minipage}{0.49\textwidth}
\includegraphics[width=\textwidth,clip=true,trim=1.6cm 0 1.2cm 0]{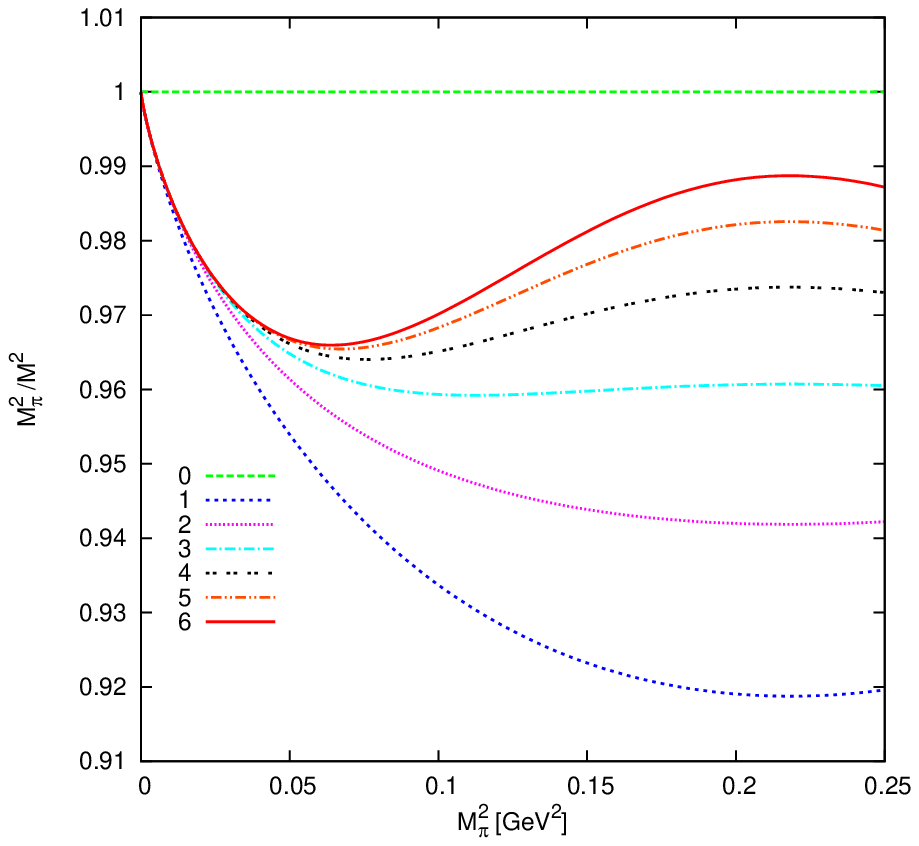}
\end{minipage}
\caption{\label{figmass1}
The contribution of the leading logarithms to $M^2_\pi/M^2$ order by order for
$F=0.090$~GeV, $F_\pi=0.0922$~GeV, $\mu =0.77$~GeV and $N=3$.
The left panel shows the expansion in $L$ keeping $F$ fixed,
the right panel the expansion in $L_{\phys}$ keeping $F_\pi$ fixed.}
\end{figure}

The coefficients  $a_i$ and $c_i$ in the expansion
of the physical mass are listed up to sixth order in
Tables~\ref{tabmass1} and~\ref{tabmass2}.
The sixth-order results are new. We can use these results
to check the expansions and how fast they converge.
We chose $F=0.090$~GeV, $F_\pi = 0.0922$~GeV and $\mu=0.77$~GeV for the plots presented here
in Fig.~\ref{figmass1}.
\begin{table}[t]
\small
\begin{center}
\begin{tabular}{|c|c|l|}
\hline
$i$ & $a_i$ for $N=3$ & $a_i$ for general $N$\\
\hline
1 & $-1/2$        & $ 1 - 1/2~N$\\[1mm]
2 & 17/8       & $7/4
          - 7/4~N
          + 5/8~N^2$\\[1mm]
3 & $-103/24$     & $ 37/12
          - 113/24~N
          + 15/4~N^2
          - N^3 $ \\[1mm]
4 & 24367/1152 & $ 839/144
          - 1601/144~N
          + 695/48~N^2
          - 135/16~N^3 
          + 231/128~N^4 $ \\[1mm]
5 & $-8821/144$   & $33661/2400
          - 1151407/43200~N
          + 197587/4320~N^2
          - 12709/300~N^3$\\
& & $
          + 6271/320~N^4
          - 7/2~N^5 $ \\[1mm]
6 & $\frac{1922964667}{6220800}$ & $  158393809/3888000
                    - 182792131/2592000~N
                    + 1046805817/7776000~N^2$\\
& & $
                    - 17241967/103680~N^3
                    + 70046633/576000~N^4
                    - 23775/512~N^5$\\
& & $
                    + 7293/1024~N^6$\\
\hline
\end{tabular}
\end{center}
\caption{\label{tabmass1} The coefficients $a_i$ of the leading logarithm $L^i$
up to $i=6$ for the physical meson mass.}
\end{table}
\begin{table}[t]
\small
\begin{center}
\begin{tabular}{|c|c|l|}
\hline
$i$ & $c_i$ for $N=3$ & $c_i$ for general $N$\\
\hline
1 & $ - 1/2$ & $ 1 - 1/2\,N$\\[1mm]
2 & $ 7/8 $  & $  - 1/4 + 3/4\,N - 1/8\,N^2 $\\[1mm]
3 & $ 211/48 $ & $  - 5/12 + 7/24\,N + 5/8\,N^2 - 1/16
         \,N^3 $\\[1mm]
4 & $ 21547/1152 $ & $ 347/144 - 587/144\,N + 47/24\,N^2
          + 25/48\,N^3 - 5/128\,N^4 $\\[1mm]
5 & $ 179341/2304 $ &$  - 6073/1800 + 32351/2400\,N -
         59933/4320\,N^2 + 224279/43200\,N^3$\\
 & & $
  + 761/1920\,N^4 - 7/256\,N^5 $\\[1mm]
6 & $ \frac{2086024177}{6220800} $ & $
-17467151/3888000 - 10487351/2592000~N
 + 68244763/1944000~N^2$\\
 & & $ - 5630053/172800~N^3 + 18673489/1728000~N^4 + 583/2560~N^5 $\\
 & & $- 21/1024~N^6
$\\
\hline
\end{tabular}
\end{center}
\caption{\label{tabmass2} The coefficients $c_i$ of the leading logarithm
$L^i_\phys$ up to $i=6$ for the physical meson mass.}
\end{table}

\subsection{Decay constant}

The decay constant $F_\pi$ is defined by
\begin{equation}
\langle 0| j^b_{a,\mu}| \phi^c(p)\rangle = i F_\pi
p_\mu\delta^{bc}\,.
\end{equation}
We thus need to evaluate a matrix-element with one external axial field
and one incoming meson. The diagrams needed for the wave function
renormalization were already evaluated in the calculation for the mass
in the previous subsection. What remains is thus the evaluation of all
relevant one-particle-irreducible (1PI) diagrams with an external $a^a_\mu$. 
At one-loop order there
is only a single diagram, but at higher orders, we must calculate
2, 4, 7, 13, 23, \ldots diagrams, not counting the auxiliary diagrams required
for the renormalization of higher-order vertices with more than two legs.
We do not show these diagrams here, but they can be found up to third order
in~\cite{Bijnens:2010xg}. The total number of diagrams with an axial
current that needs to be calculated for the decay constant to order $n$ is 
$1, 5, 18, 56, 169, 511,\ldots$

We give the coefficients for both leading logarithm series with $O_\phys = F_\pi$
and $O_0 = F$ in Tables~\ref{tab:decay} and~\ref{tab:decay2}. The sixth order is
again a new result. Note that once the expression of $F_\pi$ as a function of
$F$ is known one may express the remaining observables as a function of the
physical $M^2_\pi$ and $F_\pi$. This has already been used to calculate the
coefficients $c_i$ in Tables~\ref{tabmass2} and~\ref{tab:decay2}
from the corresponding $a_i$.
\begin{table}[t]
\small
\begin{center}
\begin{tabular}{|c|c|l|}
\hline
$i$ & $a_i$ for $N=3$ & $a_i$ for general $N$\\
\hline
1 & $ 1$ & $
          - 1/2
          + 1/2\,N
          $\\[1mm]
2 & $ - 5/4$ & $
          - 1/2
          + 7/8\,N
          - 3/8\,N^2
          $\\[1mm]
3 & $ 83/24 $ & $
          - 7/24
          + 21/16\,N
          - 73/48\,N^2
          + 1/2\,N^3
          $\\[1mm]
4 & $ - 3013/288 $ & $
           47/576
          + 1345/864\,N
          - 14077/3456\,N^2
          + 625/192\,N^3
          - 105/128\,N^4
          $\\[1mm]
5 & $  2060147/51840 $ & $
          - 23087/64800
          + 459413/172800\,N
          - 189875/20736\,N^2
    $\\ & & $
          + 546941/43200\,N^3
          - 1169/160\,N^4
          + 3/2\,N^5
          $\\[1mm]
6 & $ -\frac{69228787}{466560}  $ & $

	  - 277079063/93312000
          + 1680071029/186624000~N
$\\ & & $
          - 686641633/31104000\,N^2
          + 813791909/20736000\,N^3
$\\ & & $
          - 128643359/3456000\,N^4
          + 260399/15360\,N^5
          - 3003/1024\,N^6
$\\
\hline
\end{tabular}
\end{center}
\caption{\label{tab:decay} The coefficients $a_i$ of the leading logarithm
 $L^i$ for the decay constant $F_\pi$ in the case $N=3$  and
 in the generic $N$ case.}
\end{table}
\begin{table}[t]
\small
\begin{center}
\begin{tabular}{|c|c|l|}
\hline
$i$ & $c_i$ for $N=3$ & $c_i$ for general $N$\\
\hline
1 & $ 1 $ & $  - 1/2 + 1/2\,N $\\[1mm]
2 & $ 5/4 $ & $ 1/2 - 7/8\,N + 3/8\,N^2 $\\[1mm]
3 & $  13/12 $ & $   - 1/24 + 13/16\,N - 13/12\,N^2 + 5/
         16\,N^3 $\\[1mm]
4 & $  - 577/288$ & $  - 913/576 + 2155/864\,N - 361/
         3456\,N^2 - 69/64\,N^3
         + 35/128\,N^4 $\\[1mm]
5 & $   - 14137/810 $ & $  535901/129600 - 2279287/172800\,N
          + 273721/20736\,N^2
		$\\ & & $
     - 11559/3200\,N^3 - 997/1280\,N^4 + 63/256\,N^5 $\\[1mm]
6 & $-\frac{37737751}{466560} $  &
$
-112614143/93312000 + 3994826029/186624000\,N
$\\ & & $
-1520726023/31104000\,N^2 +276971363/6912000\,N^3
$\\ & & $
 - 39882839/3456000\,N^4 - 979/15360\,N^5
+ 231/1024\,N^6
$\\
\hline
\end{tabular}
\end{center}
\caption{\label{tab:decay2} The coefficients $c_i$ of the leading logarithm
 $L^i_\phys$ for the decay constant $F_\pi$ in the case $N=3$
and in the generic $N$ case.}
\end{table}

\begin{figure}
\begin{minipage}{0.49\textwidth}
\includegraphics[width=\textwidth,clip=true,trim=1.6cm 0 1.2cm 0]{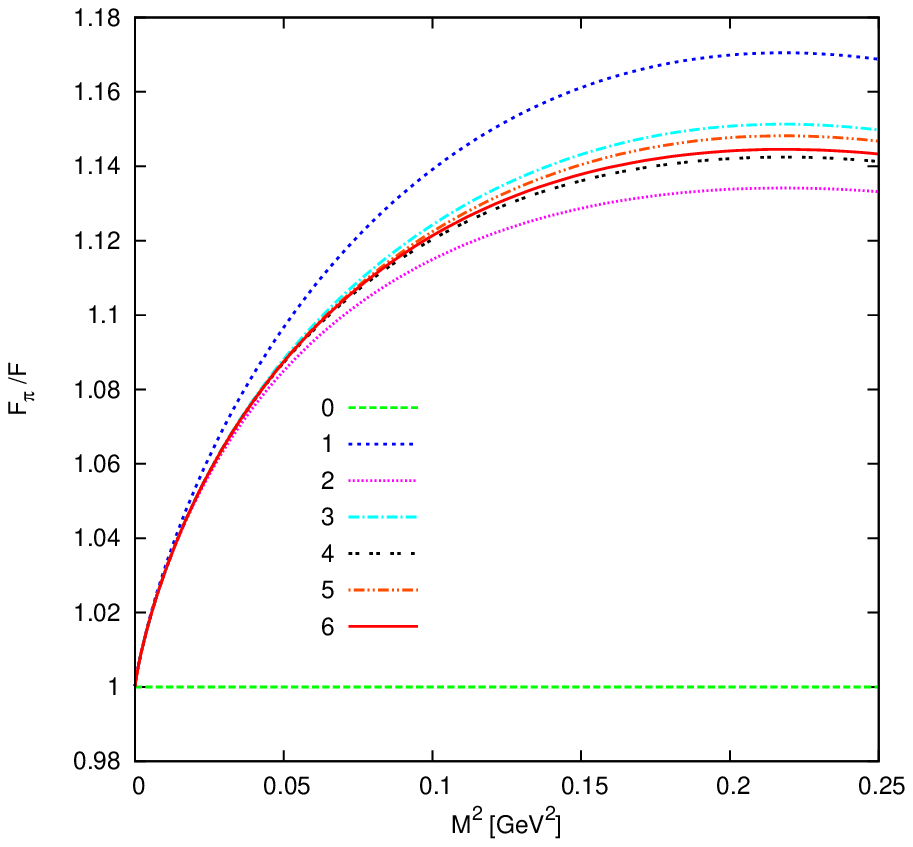}
\end{minipage}
\begin{minipage}{0.49\textwidth}
\includegraphics[width=\textwidth,clip=true,trim=1.6cm 0 1.2cm 0]{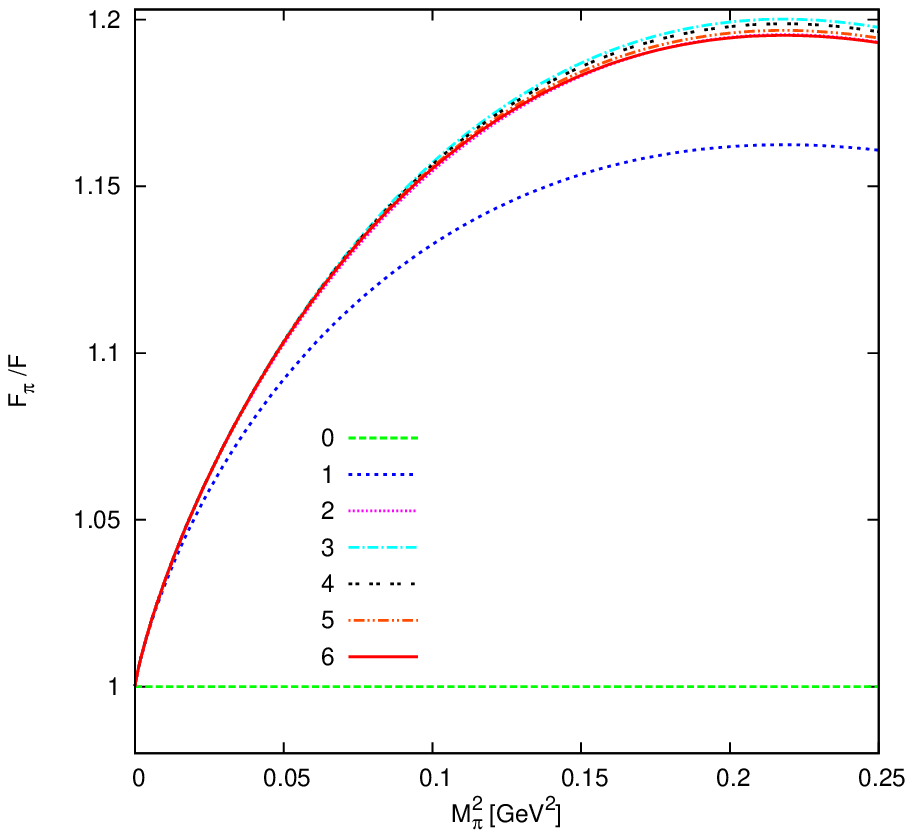}
\end{minipage}
\caption{\label{figdecay}
The contribution of the leading logarithms to $F_\pi/F$ order by order for
$\mu =0.77$~GeV and $N=3$.
The left panel shows the expansion in $L$ with $F=0.090~$GeV fixed,
the right panel the expansion in $L_\phys$ with $F_\pi=0.0922~$MeV fixed.}
\end{figure}

We have plotted in Fig.~\ref{figdecay} the expansion in terms of the
unrenormalized quantities and in terms of the physical quantities.
In both cases we get a good convergence but it is excellent for the expansion in
physical quantities.

\subsection{Vector form factor} \label{secvff}

Before we proceed to the discussion of anomalous processes, we turn to the last
ingredient from the even intrinsic parity sector
which will be used later: the vertex involving
a single photon and an even number of pions. It is directly connected to the
vector form factor, which is defined by
\begin{equation}
\langle\phi^a(p_f)|
  j^{cd}_{V,\mu}-j^{dc}_{V,\mu}|\phi^{b}(p_i)\rangle
= \left(\delta^{ac}\delta^{db}-\delta^{ad}\delta^{bc}\right)
    i(p_f+p_i)_\mu
  F_V\left[(p_f-p_i)^2\right]\,.
\end{equation}
The procedure to find the leading logarithm
for this observable follows the lines of
the one for the decay constant. For the wave function renormalization
one may again use the results obtained in the mass calculation.
We express here the results  in terms of $\tilde t= t/M_\pi^2$ and
\begin{equation}
 L_\M = \frac{M_\pi^2}{16 \pi^2 F_\pi^2} \log \frac{\mu^2}{\M^2}
\end{equation}
with a scale $\M^2$ that is some combination of $t$ and $M_\pi^2$.
Again we have added one more order compared to the result
in~\cite{Bijnens:2010xg}.
To fifth order we find
{\small
\begin{align}
\label{resultFV}
F_V&(t) = 1+ L_\mathcal{M} \Big[
           1/6\,\tilde t
          \Big]
       +  L_\mathcal{M}^2  \Big[
          \tilde t\,(- 11/12
          + 5/12\,N)
          +\tilde t^2\,( 5/36
          - 1/24\,N)
           \Big]\nonumber \\[1mm]&
       +  L_\mathcal{M}^3   \Big[
         \tilde t \,( + 1387/648
          - 845/324\,N
          + 7/9\,N^2)
         + \tilde t^2\, (- 4007/6480
\nonumber\\&
          + 3521/6480\,N
          - 29/180\,N^2)
         + \tilde t^3\, (+ 721/12960
          - 47/1440\,N
          + 1/80\,N^2)
          \Big]
\nonumber \\[1mm]&
       + L_\mathcal{M}^4   \Big[
          \tilde t \,(- 44249/15552
          + 222085/31104\,N
          - 55063/10368\,N^2
          + 127/96\,N^3)
\nonumber\\&
         + \tilde t^2 \,(+ 349403/155520
          - 15139/4860\,N
          + 86719/51840\,N^2
          - 199/480\,N^3)
\nonumber\\&
         + \tilde t^3 \,(- 85141/155520
          + 885319/1555200\,N
          - 5303/19200\,N^2
          + 21/320\,N^3)
\nonumber\\&
         +\tilde t^4\, ( + 4429/103680
          - 57451/1555200\,N
          + 289/14400\,N^2
          - 1/240\,N^3)
           \Big]
\nonumber \\[1mm]&
       + L_\mathcal{M}^5   \Big[
	\tilde t \,(
	-2278099/777600
	-2377637/466560\,N
	+64763783/4665600\,N^2
\nonumber\\&
	-178063/19200\,N^3
	+69/32\,N^4
	)
	+ \tilde t^2 \,(
	-62212433/11664000
\nonumber\\&
	+27685279/2332800\,N
	-376597697/38880000\,N^2
\nonumber\\&
	+53519593/12960000\,N^3
	-361/400\,N^4
	)
	+ \tilde t^3 \,(
	74033879/30240000
\nonumber\\&
	-2247054421/544320000\,N
	+32125153/11340000\,N^2
\nonumber\\&
	-13264877/12096000\,N^3
	+1209/5600\,N^4
	)
	+ \tilde t^4 \,(
	-299603257/816480000
\nonumber\\&
	+213192107/408240000\,N
	-98330371/272160000\,N^2
\nonumber \\&
	+546331/3780000\,N^3
	-233/8400\,N^4
	)
	+ \tilde t^5 \,(
	3090331/163296000
\nonumber \displaybreak[0] \\&
	-36097349/1632960000\,N
	+28441883/1632960000\,N^2
\nonumber \\&
	-15971/2268000\, N^3
	+1/672\,N^4
	) \Big]
\,.
\end{align}}
Note that $F_V(0) =1$ as it should be.

The vector form factor was also calculated in the massless case in
\cite{Kivel:2009az}.
In order to transform our result into this limit, we define
\begin{equation}\label{Kt}
 K_t \equiv \frac{t}{16\pi^2 F^2} \log\Big(-\frac{\mu^2}{t} \Big) \, .
\end{equation}
Replacing $\M^2 \to t$ and then performing the
limit $M_\pi^2 \to 0$ (which implies $F_\pi \to F$), we get
\begin{eqnarray}
\label{FVchiral}
\lefteqn{F_V^0(t) = 1+ K_t/6 + K_t^2 (5/36-N/24)}
\nonumber\\&&
+ K_t^3 (721/12960- 47/1440\,N + N^2/80)
\nonumber\\&&
+K_t^4 (4429/103680- 57451/1555200\,N + 289/14400\,N^2 - N^3/240 )
\nonumber\\&&
+K_t^5 (3090331/163296000 - 36097349/1632960000\,N
\nonumber\\&& \phantom{+}
+ 28441883/1632960000\,N^2- 15971/2268000\,N^3 + N^4/672) \, .
\end{eqnarray}
This differs from the result of~\cite{Kivel:2009az}. The difference
is discussed in the appendix using a dispersive approach as an alternative
check, which agrees with \cite{Bijnens:2010xg} and (\ref{FVchiral}).
Up to the given order, the coefficient of the highest power in $N$
in~(\ref{FVchiral}) at each
order is of the form
\begin{equation}
 f_V^n  =  \frac{(-1)^{n-1}}{(n+1)(n+2)} \left(\frac{N}{2}\right)^{n-1} \, ,
\end{equation}
such that, assuming this representation of $f_V^n$ to be valid
also at higher orders, we can write
\begin{equation}
 F_V^0(t) = 1 + \sum_{n=1}^\infty f_V^n K_t^n (1 + O(1/N) )\,.
\end{equation}
The summation can be performed explicitly and we obtain
the next-to-large $N$ result in the chiral limit in a closed form:
\begin{equation}
\label{FV0NLN}
F_V^{0NLN}(t) =
1+\frac1N +\frac{4}{K_tN^2} \bigg[
1-\Big(1+\frac{2}{K_tN}\Big) \log \Big(1 + \frac{K_tN}{2} \Big)
\bigg]\,.
\end{equation}
These large $N$ formulas are also
present in~\cite{Kivel:2009az} up to a sign mistake. In that article, the large $N$ has been explicitly
calculated to all orders.

\begin{table}[t]
\small
\begin{center}
\begin{tabular}{|c|c|l|}
\hline
$i$ & $c_i$ for $N=3$ & $c_i$ for general $N$\\
\hline
1  & 1          &  1\\[1mm]
2  & 2          &   $-11/2 +5/2\,N$\\[1mm]
3  & 853/108    &  $1387/108 -845/54\,N + 14/3\,N^2$ \\[1mm]
4 & 50513/1296 &  $-44249/2592 +222085/5184\,N -55063/1728\,N^2 + 127/16\, N^3$\\[1mm]
5  & 120401/648 &  $-2278099/129600 -2377637/77760\,N + 64763783/777600\,N^2 $ \\ && $- 178063/3200\,N^3 + 207/16\,N^4$\\
\hline
\end{tabular}
\end{center}
\caption{\label{tab:radius}
The coefficients $c_i$ of the leading logarithm $L_\phys$ in the expansion
of the radius $\langle r^2 \rangle_V$ in the case $N=3$ and for general $N$.}
\end{table}

\begin{table}[t]
\small
\begin{center}
\begin{tabular}{|c|c|l|}
\hline
$i$ & $c_i$ for $N=3$ & $c_i$ for general $N$\\
\hline
1  & 0         &  0\\[1mm]
2  & $1/72$          &   $5/36-1/24\,N$\\[1mm]
3 & $-71/162$    &  $ -4007/6480 +3521/6480\,N -29/180\,N^2$ \\[1mm]
4 & $-25169/7776$ &  $ 349403/155520 -15139/4860\,N + 86719/51840\,N^2 -199/480\,N^3$\\[1mm]
5 & $-1349303 /72900$ &  $ -62212433/11664000 + 27685279/2332800\,N $ \\ && $-376597697/38880000\,N^2 + 53519593/12960000\,N^3 -361/400\,N^4$\\
\hline
\end{tabular}
\end{center}
\caption{\label{tab:curvature}
The coefficients $c_i$ of the leading logarithm $L_\phys$ in the expansion
of the curvature $c_V$ in the case $N=3$ and for general $N$.}
\end{table}

We close this section with giving the expansion for the radius and curvature
of the vector form factor defined by
\begin{equation}
 F_V(t) = 1 + \frac16 \langle r^2 \rangle_V t + c_V t^2 + \ldots\,.
\end{equation}
The coefficients $c_i$ for the expansion in physical quantities are given
in Tables~\ref{tab:radius} and~\ref{tab:curvature} in units of $M_\pi^2$, again
adding one order compared to~\cite{Bijnens:2010xg}.
The result up to two-loop order agrees with the LL extracted from the full
two-loop calculation~\cite{Bijnens:1998fm}. We do not present numerical results
for the vector form factor since these are dominated in the physical
case $N=3$ by the large higher-order coefficient contributions,
see, e.g.,~\cite{Gasser:1983yg,Bijnens:1998fm}.

\section{Leading logarithms in the anomalous sector} \label{secLLodd}
\subsection{\texorpdfstring{$\pi\gamma\to\pi\pi$}{Pi gamma to pi pi}}

The process $\pi^0 \gamma \to \pi^0 \pi^0$ is forbidden by $C$-symmetry and we
will therefore concentrate on $\pi^- \gamma \to \pi^-\pi^0$.
The latter can be represented by the
anomalous \textit{VAAA} quadrangle diagram at quark level.
We follow the notation
introduced in~\cite{Bijnens:1989ff} where the one-loop order was calculated.
At tree-level, the amplitude for $\pi^-(p_1)\gamma(k)
\to \pi^- (p_2) \pi^0(p_0)$
can be obtained from the Wess-Zumino-Witten Lagrangian~(\ref{LWZW}):
\begin{equation}
A_0 = \frac{i e}{4 \pi^2 F^3}
\epsilon_{\mu\nu\alpha\beta}\varepsilon^\mu(k)p_1^\nu p_2^\alpha p_0^\beta\,.
\end{equation}
For higher orders we express the results in terms of physical
variables only as
\begin{equation}
A = iF^{3\pi}(s,t,u)
\epsilon_{\mu\nu\alpha\beta}\varepsilon^\mu(k)p_1^\nu p_2^\alpha p_0^\beta\,,
\end{equation}
with the Mandelstam variables
\begin{equation}
\label{stu}
 s = (p_1 + k)^2,\quad
 t = (p_1 -p_2)^2,\quad
 u = (p_1 -p_0)^2, \quad s+t+u = 3M_\pi^2+k^2\,.
\end{equation}
The function $F^{3\pi}(s,t,u)$ for $\gamma\pi\to\pi\pi$ is fully symmetric
in $s,t,u$. We write it in terms of $F_\pi$ as
\begin{equation}
F^{3\pi}(s,t,u) = F^{3\pi}_0 f(s,t,u)\,,\quad F^{3\pi}_0 = \frac{e}{4\pi^2F_\pi^3}\,.
\label{F3pidef}
\end{equation}

If one expands $f(s,t,u)$ as a polynomial in $s,t,u$ one can use the
relation in~(\ref{stu}) to see how many new independent kinematical
quantities can appear at each order. Up to fifth order in $s,t,u$ there
is only one at each order. At sixth order there are two. For the first five
orders we choose as independent quantities\footnote{The same arguments can
be applied to any process fully symmetric in $s$, $t$, and $u$. A prominent example is $\eta\to 3\pi^0$,
where $s+t+u=3 M_\pi^2+M_\eta^2$\,.}
\begin{equation}
\label{defDelta}
\Delta_n = s^n + t^n + u^n\,.
\end{equation}
We also define $\tilde k^2 = k^2/M_\pi^2$ and
$\tilde \Delta_n = \Delta_n/M_\pi^{2n}$.
In the end we write $f(s,t,u)$ in terms of
$\tilde k^2,M^2_\pi,\tilde\Delta_2,\ldots,\tilde\Delta_5$.

\begin{figure}
\begin{picture}(60,90)(10,-10)
\SetScale{1.0}
\SetPFont{Helvetica}{7}
\BCirc(40,40){20}
\Photon(40,20)(55.115,6.90279){2}{3}
\Line(40,20)(45.6347,0.810141)
\Line(40,20)(34.3653,0.810141)
\Line(40,20)(24.885,6.90279)
\BText(40,20){0}
\end{picture}
\begin{picture}(60,90)(10,-10)
\SetScale{1.0}
\SetPFont{Helvetica}{7}
\BCirc(40,40){20}
\Photon(40,20)(49.1645,2.22329){2}{3}
\Line(40,20)(30.8355,2.22329)
\BText(40,20){0}
\Line(40,60)(30.8355,77.7767)
\Line(40,60)(49.1645,77.7767)
\BText(40,60){0}
\end{picture}
\begin{picture}(60,90)(10,-10)
\SetScale{1.0}
\SetPFont{Helvetica}{7}
\BCirc(40,40){20}
\Photon(40,20)(55.115,6.90279){2}{3}
\Line(40,20)(45.6347,0.810141)
\Line(40,20)(34.3653,0.810141)
\Line(40,20)(24.885,6.90279)
\BText(40,20){1}
\end{picture}
\begin{picture}(60,90)(10,-10)
\SetScale{1.0}
\SetPFont{Helvetica}{7}
\BCirc(40,40){20}
\Photon(40,20)(49.1645,2.22329){2}{3}
\Line(40,20)(30.8355,2.22329)
\BText(40,20){0}
\Line(40,60)(30.8355,77.7767)
\Line(40,60)(49.1645,77.7767)
\BText(40,60){1}
\end{picture}
\begin{picture}(60,90)(10,-10)
\SetScale{1.0}
\SetPFont{Helvetica}{7}
\BCirc(40,40){20}
\Photon(40,20)(55.115,6.90279){2}{3}
\Line(40,20)(45.6347,0.810141)
\Line(40,20)(34.3653,0.810141)
\Line(40,20)(24.885,6.90279)
\BText(40,20){0}
\BText(40,60){1}
\end{picture}
\begin{picture}(60,90)(10,-10)
\SetScale{1.0}
\SetPFont{Helvetica}{7}
\BCirc(40,40){20}
\Photon(40,20)(49.1645,2.22329){2}{3}
\Line(40,20)(30.8355,2.22329)
\BText(40,20){1}
\Line(40,60)(30.8355,77.7767)
\Line(40,60)(49.1645,77.7767)
\BText(40,60){0}
\end{picture}
\begin{picture}(70,90)(5,-10)
\SetScale{1.0}
\SetPFont{Helvetica}{7}
\BCirc(40,40){20}
\Photon(40,20)(49.1645,2.22329){2}{3}
\Line(40,20)(30.8355,2.22329)
\BText(40,20){0}
\Line(22.6795,50)(2.70215,50.9516)
\Line(22.6795,50)(11.8667,66.8251)
\BText(22.6795,50){0}
\BText(57.3205,50){1}
\end{picture}
\caption{The irreducible diagrams for the process $\pi\gamma\to\pi\pi$ up to
two-loop level. The first two diagrams are the one-loop level.}
\label{figir3p}
\end{figure}
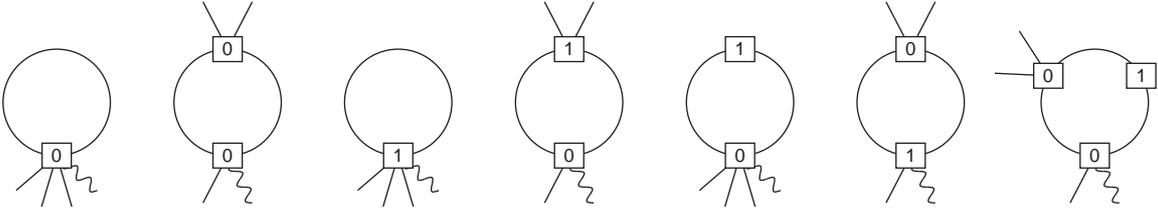

\begin{figure}
 \begin{picture}(80,90)(0,-10)
\SetScale{1.0}
\SetPFont{Helvetica}{7}
\BCirc(40,40){20}
\Photon(40,20)(57.0421,9.53264){2}{3}
\Line(40,20)(51.4901,3.62997)
\Line(40,20)(44.0518,0.414717)
\Line(40,20)(35.9482,0.414717)
\Line(40,20)(28.5099,3.62997)
\Line(40,20)(22.9579,9.53264)
\BText(40,20){0}
\end{picture}
\begin{picture}(80,90)(0,-10)
\SetScale{1.0}
\SetPFont{Helvetica}{7}
\BCirc(40,40){20}
\Photon(40,20)(49.1645,2.22329){2}{3}
\Line(40,20)(30.8355,2.22329)
\BText(40,20){0}
\Line(40,60)(24.885,73.0972)
\Line(40,60)(34.3653,79.1899)
\Line(40,60)(45.6347,79.1899)
\Line(40,60)(55.115,73.0972)
\BText(40,60){0}
\end{picture}
\begin{picture}(80,90)(0,-10)
\SetScale{1.0}
\SetPFont{Helvetica}{7}
\BCirc(40,40){20}
\Photon(40,20)(55.115,6.90279){2}{3}
\Line(40,20)(45.6347,0.810141)
\Line(40,20)(34.3653,0.810141)
\Line(40,20)(24.885,6.90279)
\BText(40,20){0}
\Line(40,60)(30.8355,77.7767)
\Line(40,60)(49.1645,77.7767)
\BText(40,60){0}
\end{picture}
\begin{picture}(80,90)(0,-10)
\SetScale{1.0}
\SetPFont{Helvetica}{7}
\BCirc(40,40){20}
\Photon(40,20)(49.1645,2.22329){2}{3}
\Line(40,20)(30.8355,2.22329)
\BText(40,20){0}
\Line(22.6795,50)(2.70215,50.9516)
\Line(22.6795,50)(11.8667,66.8251)
\BText(22.6795,50){0}
\Line(57.3205,50)(68.1333,66.8251)
\Line(57.3205,50)(77.2979,50.9516)
\BText(57.3205,50){0}
\end{picture}
\caption{The irreducible (auxiliary) diagrams needed for the vertex
$5\pi\gamma$ up to one-loop level.}
\label{figir5p}
\end{figure}
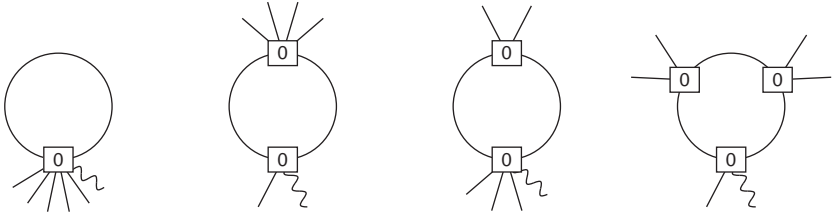

The main focus of this article is calculating the leading logarithms
in the anomalous sector.
The procedure follows very similar steps as in the
even sector. 
From the even sector we will need the wave function renormalization
and the expressions for the higher-order purely mesonic vertices
as well as the decay constant and mass 
when using an expansion in terms of physical quantities.
Vertices coupling an even number of pions to a single photon are not
needed at this point, because the two-flavour anomaly
does not contain interaction
terms involving an odd number of pions and no photon.
What remains to be calculated are the irreducible
diagrams with three external pions and one external photon. 
The required diagrams up to two-loop order are depicted in Fig.~\ref{figir3p}.
As in Section~\ref{sec:LLeven},
a box with $n$ inside,~%
\begin{picture}(11,0)(0,0)
\SetScale{1.0}
\SetPFont{Helvetica}{7}
\BText(5,4){n}
\end{picture}, 
means a vertex of order $n$.
To reach the one-loop level, we must calculate the first two of these diagrams.
Inspection of the remaining diagrams for the two-loop level
shows that all vertices are already known except
the one with five pions and one photon in the third diagram.
The diagrams that are needed in order to obtain
its divergence are shown in Fig.~\ref{figir5p}. Note that the vertex
with three pions and one photon in the sixth diagram of Fig.~\ref{figir3p} is
fixed by the one-loop calculation, the first two diagrams in the same figure.

To go to higher orders we rapidly need vertices with many more legs.
We have generated all diagrams needed and calculated them up to fifth order.
We agree with the logarithm determined from the
full one-loop result~\cite{Bijnens:1989ff}. The results were obtained in several
different parametrizations with consistent results.

We have calculated the amplitude for $\pi^-\gamma\to\pi^-\pi^0$ up to five-loop
level. In the general case, i.e., for $k^2,M_\pi^2\ne 0$, we obtain
\begin{eqnarray}
\lefteqn{f^{LL}(s,t,u) = 1
+ L_\mathcal{M}   \frac{1}{6} \left(3+\tilde k^2\right) 
+ L_\mathcal{M}^2 \frac{1}{72} (\tilde k^2-3) (\tilde k^2+33) }&&
\nonumber\\&&
+ L_\mathcal{M}^3 \frac{1}{1296}\big(90 \tilde\Delta_3 -640\tilde\Delta_2
 - 8157 + 2105 \tilde k^2 + 81 \tilde k^4 + \tilde k^6\Big)
+ L_\mathcal{M}^4  \frac{1}{155520}\Big[
        -1532\tilde\Delta_4
\nonumber\\&&
       +\tilde\Delta_3 (   88538 + 1890\tilde k^2 )
       - \tilde\Delta_2 ( 577760 + 12240\tilde k^2 + 540 \tilde k^4 )
       - 2433375 + 1296190 \tilde k^2 
\nonumber\\&&
       + 57430 \tilde k^4 + 480 \tilde k^6 
       + 185 \tilde k^8\Big]
+ L_\mathcal{M}^5 \frac{1}{326592000}\Big[
        \tilde\Delta_5 (   13252156 )
\nonumber\\&&
       - \tilde\Delta_4 ( 160744570 + 518350 \tilde k^2 )
       + \tilde\Delta_3 (  1465187530 + 39593272 \tilde k^2 
                         + 247260 \tilde k^4 )
\nonumber\\&&
       - \tilde\Delta_2 ( 6756522937 + 257781206 \tilde k^2 
            + 11188776 \tilde k^4 - 9160 \tilde k^6 )
       - 6498695163 
\nonumber\\&&
       + 12675091794 \tilde k^2 + 801259373 \tilde k^4 
       + 4780240 \tilde k^6 + 2948600 \tilde k^8 - 1832 \tilde k^{10}\Big]\,.
\label{fLL}
\end{eqnarray}
The symmetry in $s$, $t$ and $u$ is obvious in this expression.
Note that at second order $\tilde\Delta_2$ does not appear even though it
could be present a priori.

We can check whether we find a simpler expression in the massless
limit and for an on-shell photon, $k^2=0$. In this case we need to express
the result in terms of the logarithm
\begin{equation}
L_\Delta = \frac{1}{16\pi^2 F^2}\log\left(\frac{\mu^2}{\hat\Delta}\right)\,,
\end{equation}
where $\hat\Delta$ is some combination of $s$, $t$, and $u$:
\begin{equation}
 f^{LL\,0}(s,t,u) =
 1 +\frac{5}{72} \Delta_3 L_\Delta^3
- \frac{383}{19440} \Delta_4 L_\Delta^4
+ \frac{3313039}{81648000} \Delta_5 L_\Delta^5\,.
\end{equation}
It is clear from symmetry considerations that there should be no term
linear in $L_\Delta$.
We do however not know why the quadratic term is also absent.

\subsection{\texorpdfstring{$\pi^0\to\gamma\gamma$}{Pi0 to gamma gamma}}

This is the most important process in the odd-intrinsic
parity sector of QCD, since it is the process
in which the chiral anomaly was discovered. It remains the main
experimental test thereof.

We started our discussion of anomalous processes with $\pi\gamma\to\pi\pi$
because the leading logarithms for this process could be calculated
from our results in the even intrinsic parity sector and just
one type of anomalous vertex.

We define the
reduced amplitude $F_{\pi\gamma\gamma}$ for $\pi^0 \to \gamma(k_1)\gamma(k_2)$
\begin{equation}
 A = \epsilon_{\mu\nu\alpha\beta}\,
\varepsilon_1^{*\mu}(k_1)\varepsilon_2^{*\nu}(k_2)\,k_1^\alpha k_2^\beta \, 
F_{\pi\gamma\gamma}(k_1^2,k_2^2)\,.
\end{equation}
$F_{\pi\gamma\gamma}(k_1^2,k_2^2)$ is symmetric under the interchange of the two photons.

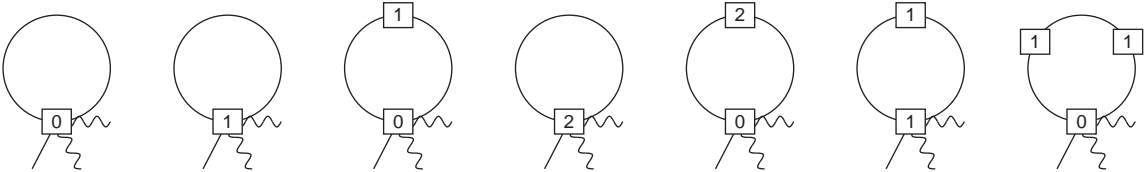
\begin{figure}[t]
\begin{picture}(60,90)(10,-10)
\SetScale{1.0}
\SetPFont{Helvetica}{7}
\BCirc(40,40){20}
\Photon(40,20)(49.1645,2.22329){2}{3}
\Photon(40,20)(60,20){2}{3}
\Line(40,20)(30.8355,2.22329)
\BText(40,20){0}
\end{picture}
\begin{picture}(60,90)(10,-10)
\SetScale{1.0}
\SetPFont{Helvetica}{7}
\BCirc(40,40){20}
\Photon(40,20)(49.1645,2.22329){2}{3}
\Photon(40,20)(60,20){2}{3}
\Line(40,20)(30.8355,2.22329)
\BText(40,20){1}
\end{picture}
\begin{picture}(60,90)(10,-10)
\SetScale{1.0}
\SetPFont{Helvetica}{7}
\BCirc(40,40){20}
\Photon(40,20)(49.1645,2.22329){2}{3}
\Photon(40,20)(60,20){2}{3}
\Line(40,20)(30.8355,2.22329)
\BText(40,20){0}
\BText(40,60){1}
\end{picture}
\begin{picture}(60,90)(10,-10)
\SetScale{1.0}
\SetPFont{Helvetica}{7}
\BCirc(40,40){20}
\Photon(40,20)(49.1645,2.22329){2}{3}
\Photon(40,20)(60,20){2}{3}
\Line(40,20)(30.8355,2.22329)
\BText(40,20){2}
\end{picture}
\begin{picture}(60,90)(10,-10)
\SetScale{1.0}
\SetPFont{Helvetica}{7}
\BCirc(40,40){20}
\Photon(40,20)(49.1645,2.22329){2}{3}
\Photon(40,20)(60,20){2}{3}
\Line(40,20)(30.8355,2.22329)
\BText(40,20){0}
\BText(40,60){2}
\end{picture}
\begin{picture}(60,90)(10,-10)
\SetScale{1.0}
\SetPFont{Helvetica}{7}
\BCirc(40,40){20}
\Photon(40,20)(49.1645,2.22329){2}{3}
\Photon(40,20)(60,20){2}{3}
\Line(40,20)(30.8355,2.22329)
\BText(40,20){1}
\BText(40,60){1}
\end{picture}
\begin{picture}(60,90)(10,-10)
\SetScale{1.0}
\SetPFont{Helvetica}{7}
\BCirc(40,40){20}
\Photon(40,20)(49.1645,2.22329){2}{3}
\Photon(40,20)(60,20){2}{3}
\Line(40,20)(30.8355,2.22329)
\BText(40,20){0}
\BText(22.6795,50){1}
\BText(57.3205,50){1}
\end{picture}
\caption{First type of the irreducible diagrams contributed to
$\pi^0\to\gamma\gamma$ up to three-loop order}
\label{figtype1}
\end{figure}

\begin{figure}[t]
\begin{picture}(70,90)(5,-10)
\SetScale{1.0}
\SetPFont{Helvetica}{7}
\BCirc(40,40){20}
\Photon(40,20)(49.1645,2.22329){2}{3}
\Line(40,20)(30.8355,2.22329)
\BText(40,20){0}
\Photon(40,60)(40,80){2}{3}
\BText(40,60){0}
\end{picture}
\begin{picture}(70,90)(5,-10)
\SetScale{1.0}
\SetPFont{Helvetica}{7}
\BCirc(40,40){20}
\Photon(40,20)(49.1645,2.22329){2}{3}
\Line(40,20)(30.8355,2.22329)
\BText(40,20){0}
\Photon(40,60)(40,80){2}{3}
\BText(40,60){1}
\end{picture}
\begin{picture}(70,90)(5,-10)
\SetScale{1.0}
\SetPFont{Helvetica}{7}
\BCirc(40,40){20}
\Photon(40,20)(49.1645,2.22329){2}{3}
\Line(40,20)(30.8355,2.22329)
\BText(40,20){1}
\Photon(40,60)(40,80){2}{3}
\BText(40,60){0}
\end{picture}
\begin{picture}(70,90)(5,-10)
\SetScale{1.0}
\SetPFont{Helvetica}{7}
\BCirc(40,40){20}
\Photon(40,20)(49.1645,2.22329){2}{3}
\Line(40,20)(30.8355,2.22329)
\BText(40,20){0}
\Photon(22.6795,50)(5.35898,60){2}{3}
\BText(22.6795,50){0}
\BText(57.3205,50){1}
\end{picture}
\begin{picture}(70,90)(5,-10)
\SetScale{1.0}
\SetPFont{Helvetica}{7}
\BCirc(40,40){20}
\Photon(40,20)(49.1645,2.22329){2}{3}
\Line(40,20)(30.8355,2.22329)
\BText(40,20){0}
\Photon(40,60)(40,80){2}{3}
\BText(40,60){2}
\end{picture}
\begin{picture}(70,90)(5,-10)
\SetScale{1.0}
\SetPFont{Helvetica}{7}
\BCirc(40,40){20}
\Photon(40,20)(49.1645,2.22329){2}{3}
\Line(40,20)(30.8355,2.22329)
\BText(40,20){1}
\Photon(40,60)(40,80){2}{3}
\BText(40,60){1}
\end{picture}

\begin{picture}(70,90)(5,-10)
\SetScale{1.0}
\SetPFont{Helvetica}{7}
\BCirc(40,40){20}
\Photon(40,20)(49.1645,2.22329){2}{3}
\Line(40,20)(30.8355,2.22329)
\BText(40,20){2}
\Photon(40,60)(40,80){2}{3}
\BText(40,60){0}
\end{picture}
\begin{picture}(70,90)(5,-10)
\SetScale{1.0}
\SetPFont{Helvetica}{7}
\BCirc(40,40){20}
\Photon(40,20)(49.1645,2.22329){2}{3}
\Line(40,20)(30.8355,2.22329)
\BText(40,20){0}
\Photon(22.6795,50)(5.35898,60){2}{3}
\BText(22.6795,50){0}
\BText(57.3205,50){2}
\end{picture}
\begin{picture}(70,90)(5,-10)
\SetScale{1.0}
\SetPFont{Helvetica}{7}
\BCirc(40,40){20}
\Photon(40,20)(49.1645,2.22329){2}{3}
\Line(40,20)(30.8355,2.22329)
\BText(40,20){0}
\Photon(22.6795,50)(5.35898,60){2}{3}
\BText(22.6795,50){1}
\BText(57.3205,50){1}
\end{picture}
\begin{picture}(70,90)(5,-10)
\SetScale{1.0}
\SetPFont{Helvetica}{7}
\BCirc(40,40){20}
\Photon(40,20)(49.1645,2.22329){2}{3}
\Line(40,20)(30.8355,2.22329)
\BText(40,20){1}
\Photon(22.6795,50)(5.35898,60){2}{3}
\BText(22.6795,50){0}
\BText(57.3205,50){1}
\end{picture}
\begin{picture}(70,90)(5,-10)
\SetScale{1.0}
\SetPFont{Helvetica}{7}
\BCirc(40,40){20}
\Photon(40,20)(49.1645,2.22329){2}{3}
\Line(40,20)(30.8355,2.22329)
\BText(40,20){0}
\Photon(20,40)(0,40){2}{3}
\BText(20,40){0}
\BText(40,60){1}
\BText(60,40){1}
\end{picture}
\begin{picture}(70,90)(5,-10)
\SetScale{1.0}
\SetPFont{Helvetica}{7}
\BCirc(40,40){20}
\Photon(40,20)(49.1645,2.22329){2}{3}
\Line(40,20)(30.8355,2.22329)
\BText(40,20){0}
\BText(20,40){1}
\Photon(40,60)(40,80){2}{3}
\BText(40,60){0}
\BText(60,40){1}
\end{picture}
\caption{Second type of the irreducible diagrams contributed to
$\pi^0\to\gamma\gamma$ up to three-loop order}
\label{figtype2}
\end{figure}
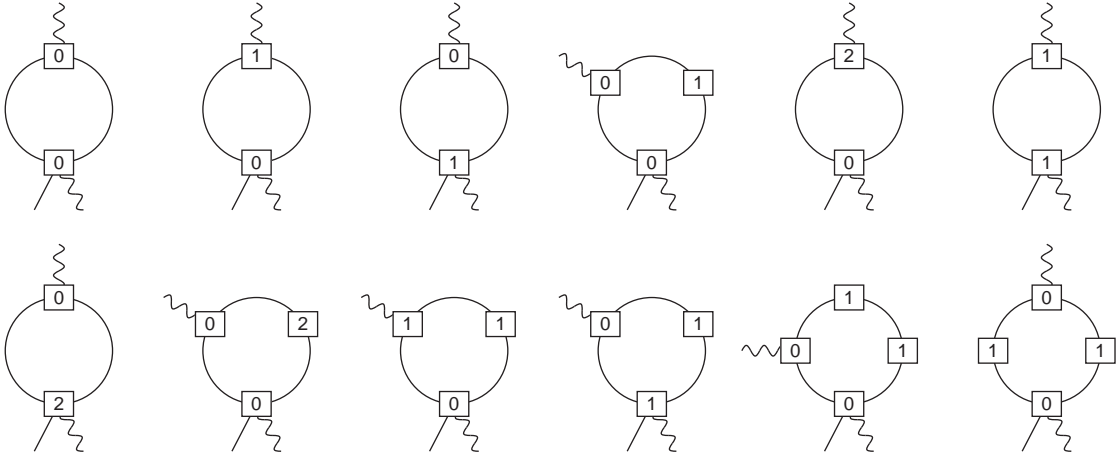

The irreducible diagrams contributing to
$\pi^0\to\gamma\gamma$ consist of two different types of one-loop diagrams.
The first type contains the diagrams where both
photons are attached to the \emph{same} vertex, while in the diagrams of
the second type, the two photons connect to two different vertices, only one
of which is anomalous.
We show here the diagrams needed for the calculation of the
leading logarithms up to third order.
Those of the first type are depicted in Fig.~\ref{figtype1}.
There is one diagram at one-loop order, there are two at two-loop order, and four
at three-loop order.
The diagrams of the second type are depicted in Fig.~\ref{figtype2}.
This time,
there is one diagram at one-loop order, there are three at two-loop, 
and eight at three-loop order.
The figures do not contain the auxiliary diagrams that are needed in order to determine
the higher-order vertices with more than one pion leg. We only mention that up
to three-loop order, one needs to calculate 11 and 23 diagrams for the first and
second type of topologies, respectively.

The first type of diagrams is a consistent subset if we only keep the
terms with two vector sources in the Lagrangian~(\ref{LWZW}). That this leads to
identical results for several different parametrizations provides a thorough
check on the correctness of our programs for this class of diagrams separately.

We write the result with $\tilde k_i^2 = k_i^2/M_\pi^2$ in the form
\begin{equation}\label{Fpigg}
F_{\pi\gamma\gamma}(k_1^2,k_2^2) = \frac{e^2}{4\pi^2 F_\pi} F_\gamma(k_1^2) F_\gamma(k_2^2)
 F_{\gamma\gamma}(k_1^2,k_2^2) \hat F\,.
\end{equation}
The functions $F_\gamma(k_i^2)$ and $F_{\gamma\gamma}(k_1^2,k_2^2)$
are defined to be equal to one for $k_i^2=0$.
$F_{\gamma\gamma}(k_1^2,k_2^2)$ contains only those parts that cannot
be absorbed in the $F_\gamma$ and thus gives the part that cannot be obtained
as a product of single photon form factors. Finally $\hat F$ gives the
corrections for the on-shell decay $\pi^0\to\gamma\gamma$.

We have calculated all contributions needed for the LLs up to six-loop order and obtained
\begin{align}
\hat F =\; &1-1/6\,L_\mathcal{M}^2+ 5/6\,L_\mathcal{M}^3
       + 56147/7776\,L_\mathcal{M}^4
       + 446502199/11664000\,L_\mathcal{M}^5
\nonumber\\&
       +65694012997/367416000\,L_\mathcal{M}^6\,,
\nonumber\\
F_\gamma(k^2) =\;
 &1+L_\mathcal{M} (1/6\,\tilde k^2)
  +L_\mathcal{M}^2 ( 5/24\,\tilde k^2 + 1/72\,\tilde k^4)
+ L_\mathcal{M}^3 (71/432\,\tilde k^2 + 1/24\,\tilde k^4
\nonumber\\&
   + 1/1296\,\tilde k^6)
       +L_\mathcal{M}^4 (- 24353/31104\,\tilde k^2 + 4873/10368\,\tilde k^4 - 2357/31104\,\tilde k^6
\nonumber\\&
          + 145/31104\,\tilde k^8)
       +L_\mathcal{M}^5(- 548440741/81648000\,\tilde k^2
          + 9793363/3024000\,\tilde k^4
\nonumber\\&
          - 32952389/54432000\,\tilde k^6
          + 487493/13608000\,\tilde k^8
          - 2069/10886400\,\tilde k^{10})\,
\nonumber\\&
		+L_\mathcal{M}^6(
			-3519465627493/102876480000\,\tilde k^2
\nonumber\\&
			+3560724235307/205752960000\,\tilde k^4
			-1524042680197/411505920000\,\tilde k^6
\nonumber\\&
			+4741599089/11757312000\,\tilde k^8
			-510932327/13716864000\,\tilde k^{10}
\nonumber\\&
			+1775869/914457600\,\tilde k^{12})\, ,
\nonumber\\
F_{\gamma\gamma}(k_1^2,k_2^2)=\;
&1+L_\mathcal{M}^3 \tilde k_1^2 \tilde k_2^2 \frac{1}{72}
+L_\mathcal{M}^4  \tilde k_1^2 \tilde k_2^2 \big[- 203/7776
                  + 29/10368\, (\tilde k_1^2+\tilde k_2^2)
\nonumber\\&
                  + 1/216\, (\tilde k_1^4+\tilde k_2^4)
                  -1/144\, \tilde k_1^2 \tilde k_2^2\big]
+L_\mathcal{M}^5 \tilde k_1^2 \tilde k_2^2 \big[- 5983633/10206000
\nonumber\\&
                + 46103/1632960\, (\tilde k_1^2+\tilde k_2^2)
                + 372113/11664000\, (\tilde k_1^4+\tilde k_2^4)
\nonumber\\&
                - 211/5443200\, (\tilde k_1^6+\tilde k_2^6)
                - 394157/9072000\, \tilde k_1^2 \tilde k_2^2
          - 4/25515\, \tilde k_1^2 \tilde k_2^2 (\tilde k_1^2+\tilde k_2^2) \big]
\nonumber \displaybreak[0] \\&
+L_\mathcal{M}^6 \tilde k_1^2 \tilde k_2^2 \big[ -1072421939773/205752960000
\nonumber \displaybreak[0] \\&
					+1444445383/6531840000\, (\tilde k_1^2+\tilde k_2^2)
					+10840553807/102876480000\, (\tilde k_1^4+\tilde k_2^4)
\nonumber\\&
					+282016297/205752960000\, (\tilde k_1^6+\tilde k_2^6)
					+6157391/4115059200\, (\tilde k_1^8+\tilde k_2^8)
\nonumber\\&
					-3852620057/29393280000\, \tilde k_1^2 \tilde k_2^2
					-154739/58320000\, \tilde k_1^2 \tilde k_2^2 (\tilde k_1^2+\tilde k_2^2)
\nonumber\\&
					-75041473/20575296000\, \tilde k_1^2 \tilde k_2^2 (\tilde k_1^4+\tilde k_2^4)
					+174329/35721000\, \tilde k_1^4 \tilde k_2^4 \big]\,.
\label{Fgamma}
\end{align}

The absence of the linear term in $\hat F$ agrees with the statement from~\cite{Donoghue:1986wv,Bijnens:1988kx}
that the contribution from one-loop diagrams at NLO can be absorbed into $F_\pi$. The quadratic term also
coincides with the two-loop calculation of~\cite{Kampf:2009tk} and the complete one-loop expression for off-shell
photons is the same as in~\cite{Bijnens:1988kx}.

Note that the nonfactorizable contribution $F_{\gamma\gamma}$
only starts at three-loop order and that the part surviving in
the chiral limit only starts at four-loop level.
The leading logarithms thus predict this part to be fairly small.

The lowest-order results for the two anomalous processes are connected via the
current algebra relation~\cite{Adler:1971nq,Terentev:1971kt}
\begin{equation}
F^{3\pi}(0,0,0) = \frac{1}{e F_\pi^2} F_{\pi\gamma\gamma}(0,0) \; ,
\end{equation}
which holds exactly in the chiral limit. Even beyond this limit it is valid also at
the one-loop level for the leading logarithms as can be
seen by comparing~(\ref{fLL}) and~(\ref{Fgamma}). The current algebra relation remains true, if in
both amplitudes one of the photons is allowed to be off-shell, i.e.
\begin{equation}
F^{3\pi}(s,t,u)\Big|_{s+t+u = 3 M_\pi^2 + k^2} = \frac{1}{e F_\pi^2} F_{\pi\gamma\gamma}(k^2,0) \; .
\end{equation}
It turns out that in the chiral limit, this relation holds for the leading logarithms up to two loops,
as can again be checked from~(\ref{fLL}) and~(\ref{Fgamma}).


\section{Phenomenology of the anomalous sector}
\label{secpheno}

\subsection{\texorpdfstring{$\pi\gamma\to\pi\pi$}{Pi gamma to pi pi}}

The only direct measurement of the $\pi\gamma\to\pi\pi$ vertex has been
performed at the IHEP
accelerator in Serpukhov~\cite{Antipov:1986tp}
using $\pi^-\gamma\to\pi^-\pi^0$, where the $\gamma$ comes from the 
electromagnetic field of a nucleus via the Primakoff effect.
The relevant cross-section was measured with a pion beam of $E=40~\text{GeV}$
and the photons' virtuality was in the region $k^2<2\times 10^{-3}~\text{GeV}^2$,
which can be neglected at the present precision.
The analysis is for values of $s< 10 M_\pi^2$ and thus within the region where ChPT
is applicable. Assuming the function
$F^{3\pi}(s,t,u)$ of \eqref{F3pidef} to be approximately constant,
$F^{3\pi}(s,t,u) \approx \bar{F}^{3\pi}$, they find
\begin{equation}
\label{F3pibar}
\bar{F}^{3\pi}_\text{exp} = 12.9\pm0.9\pm0.5~\text{GeV}^{-3} \;.
\end{equation}
Another derivation used the
data on $\pi^- e^- \to \pi^- e^- \pi^0$ \cite{Amendolia:1985bs}
to determine $F^{3\pi}_0$ \cite{Giller:2005uy}. They obtained
\begin{equation}
F^{3\pi}_\text{0,exp} = 9.9\pm1.1~\text{GeV}^{-3} \quad\text{or}\quad 9.6\pm1.1~\text{GeV}^{-3}\;,
\end{equation}
depending on the way electromagnetic corrections are included.

The value in (\ref{F3pibar}) needs to be corrected for
extrapolation to the point $s=t=u=0$.
The one-loop ChPT corrections were evaluated in \cite{Bijnens:1989ff}
and a possibly large electromagnetic correction identified
in \cite{Ametller:2001yk}. The main conclusion is that the experimental
values are in fairly good agreement with the theoretical result
of (\ref{F3pidef}) which gives $F^{3\pi}_0 = 9.8~\text{GeV}^{-3}$ as discussed
further below.

The comparison with \cite{Antipov:1986tp} goes via their result
 $\sigma/Z^2 = 1.63\pm0.23\pm0.13$~nb and~\cite{Antipov:1986tp,Ametller:2001yk}
\begin{eqnarray}
\label{sigma}
\frac{\sigma}{Z^2} &=&
\frac{\alpha}{\pi}\int_{4 M_\pi^2}^{10 M_\pi^2}  ds 
 \left(\ln\frac{q^2_\text{max}}{q^2_\text{min}}+\frac{q^2_\text{min}}{q^2_\text{max}}-1\right)
\frac{\sigma_{\gamma\pi\to\pi\pi}}{s-M_\pi^2}\,,
\nonumber\\[2mm]
\sigma_{\gamma\pi\to\pi\pi}&=&\frac{1}{1024\pi}
s(s-M_\pi^2)\left(1-\frac{4M_\pi^2}{s}\right)^{3/2}
\int_0^\pi \! d\theta \sin^3\theta \left|F_0^{3\pi}f(s,t,u)\right|^2\,.
\end{eqnarray}
Inserting the charged pion mass, $q^2_\text{max}=2 \cdot 10^{-3}~\mathrm{GeV}^{2}$,
$q^2_\text{min}= \left((s-M_\pi^2)/(2E)\right)^2$, and $f(s,t,u)=1$ in (\ref{sigma})
leads to the value in (\ref{F3pibar}).

The various known higher-order corrections can now be included via $f(s,t,u)$:
\begin{equation}
f(s,t,u) = 1 + f_0^\text{EM} + f_1^\text{loop} + f_1^\text{tree} + \ldots \,.
\end{equation}
The dependence on $s,t,u$ is tacitly assumed for all functions $f_i$. 
The index $i$ refers to the $\hbar$ order.
The leading-order electromagnetic correction $f_0^\text{EM}$ was determined
in~\cite{Ametller:2001yk} as $f_0^\text{EM} = -2 e^2 F_\pi^2/t$ and higher-order
electromagnetic corrections were found to be small.
The next-to-leading-order correction coming from one-loop graphs is,
in the isospin limit, given by~\cite{Bijnens:1989ff}
\begin{equation}\label{f1loop}
 f_1^\text{loop} = \frac{1}{6 F_\pi^2} \bigg[ -\frac{M_\pi^2}{(4\pi)^2} \bigg(1
+ 3
\log\frac{M_\pi^2}{\mu^2} \bigg) + I(s)+I(t)+I(u)
 \bigg] \,,
\end{equation}
with $I(s) = (s-4 M_\pi^2)\bar J(s)$, where
$\bar J(s)$ is the standard subtracted two-point function
\begin{equation}
16\pi^2\bar J(q^2) = \sigma \log \frac{\sigma-1}{\sigma+1} + 2\;,\qquad \sigma =
\sqrt{1-4 M_\pi^2/q^2} \;.
\end{equation}
The logarithmic term in (\ref{f1loop}) agrees with our LL calculation. 
The full expression accounting for the pion mass difference can be found
in~\cite{Ametller:2001yk}.

The contribution from the NLO Lagrangian can be expressed in terms
of the low-energy constants introduced in~\cite{Bijnens:2001bb} as
\begin{align}
 f_1^\text{tree} = 128\pi^2 M_\pi^2\, (c_2^{Wr} + c_6^{Wr})\,.
\end{align}
In order to estimate the value of $f_1^\text{tree}$, several methods
exist in the literature: hidden local symmetry (HLS)~\cite{Bijnens:1989ff},
phenomenology~\cite{Strandberg:2003zf}, the constituent quark
model (CQM)~\cite{Bijnens:1991db,Strandberg:2003zf}, Schwinger-Dyson equation
(SDE)~\cite{Jiang:2010wa}, or resonance saturation~\cite{Kampf:2011ty},
to name a few.
The spread in results can be seen from the estimates following from three
of these methods
\begin{equation}
\label{f1ctmodels}
 f_1^\text{tree} = \frac{3 M_\pi^2}{2 M_\rho^2} = 0.048 ~\text{(HLS)},
\qquad
= 0.19 ~\text{(CQM)},\qquad
= -0.01 ~\text{(SDE)}\,.
\end{equation}
\begin{table}
\begin{center}
\begin{tabular}{|c|c|c|c|c|c|c|c|}
\hline
 &LO & $f_0^\text{EM}$ & $f_1^\text{loop}$(LL) & $f_1^\text{loop}$ & $f_1^\text{tree}$ (HLS) & $f_1^\text{tree}$ (CQM) & $f_1^\text{tree}$ (SDE)\\\hline
$F_0^{3\pi}$& 12.9 & 12.3 & 12.0    & 11.9 & 11.4 & 10.1 & 12.0  \\ \hline
\end{tabular}
\end{center}
\caption{The extraction of the anomalous $\gamma3\pi$ factor
$F^{3\pi}_0$ (in GeV$^{-3}$) from experiment using
various estimates of the higher-order corrections.
LO includes no higher-order corrections and thus coincides with~(\ref{F3pibar}).
The next column contains the QED correction $f_0^\text{EM}$. 
Then come the values including in addition
the leading logarithm and the complete correction from $f_1^\text{loop}$.
The last three columns also contain $f_1^\text{tree}$ using the model
estimates in~(\ref{f1ctmodels}).}
\label{tabexpf3pi}
\end{table}
The two-loop corrections, estimated using dispersive techniques
\cite{Hannah:2001ee}, are found to be small.

These corrections can now be
incorporated in the calculation of $F^{3\pi}_0$ from the cross section
measured at Serpukhov using (\ref{sigma}).
Turning them on one by one, 
we find the results listed in Table~\ref{tabexpf3pi}.
Comparison of the third and fourth column shows that at the one-loop order,
the leading logarithm
provides a good estimate for the size of the complete correction:
it accounts for 60\% of the shift.
The uncertainty on the listed values has two main sources:
the experimental
uncertainty of about  $\pm 1~\GeV^{-3}$
and the model dependence of the $c_i^{Wr}$.
From the spread of the estimates in Table~\ref{tabexpf3pi}, the latter also is
about $\pm 1~\GeV^{-3}$.
The total error, adding quadratically, is thus about
$\pm 1.5~\GeV^{-3}$, such that the theoretical result
$ F^{3\pi}_0 = 9.8~\GeV^{-3}$ agrees reasonably with the final
values at the one-loop level.

Let us now return to the discussion of the expansion $f^{LL}$ in~\eqref{fLL}.
The one-loop LL shifts the result by $-0.3$ as already shown.
Adding the LL up contributions up to five-loop order in (\ref{sigma})
leads to
\begin{equation}
F_0^{3\pi LL} = (12.9-0.3+0.04+0.02+0.006+0.001+\ldots) \; \GeV^{-3} \,.
\end{equation}
Clearly, the series converges rather well. The small size
of the LLs beyond one loop indicates that the full corrections 
at higher orders are negligible.

\begin{figure}[t]
\begin{center}
\includegraphics[angle=270,width=0.7\textwidth]{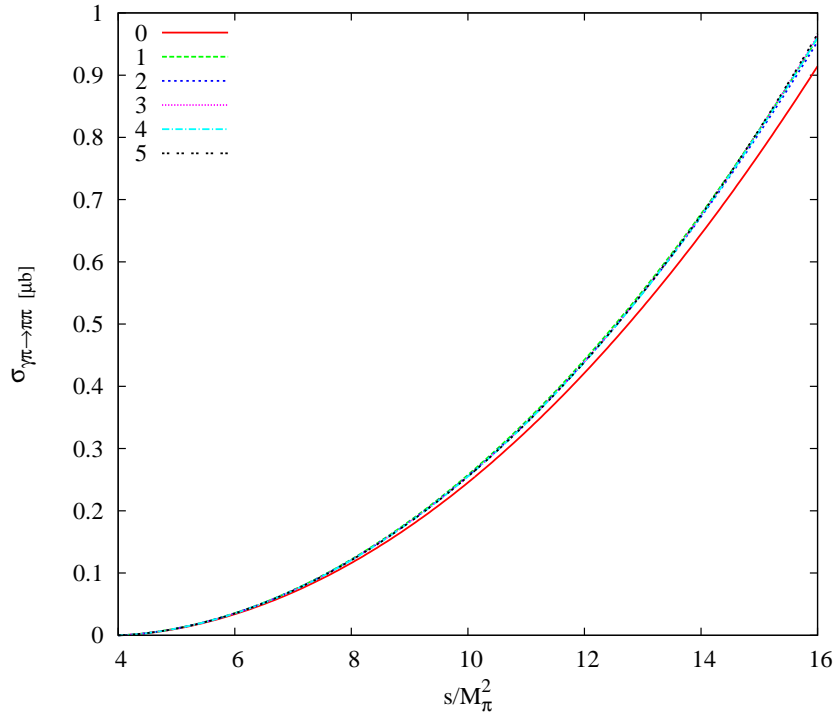}
\end{center}
\caption{\label{figf3pi}
The leading logarithm contribution to the total cross section for
$\pi^-\gamma\to\pi^-\pi^0$ as a function of $s$.}
\end{figure}

The total cross section obtained from only the
LL contributions as a function of the center-of-mass energy is depicted in Fig.~\ref{figf3pi}.

\subsection{\texorpdfstring{$\pi^0\to\gamma\gamma$}{Pi0 to gamma gamma}}

For the decay $\pi^0\to\gamma\gamma$, there is more experimental information
available. For a recent review, see \cite{Bernstein:2011bx}.
The current PDG average (\cite{Nakamura:2010zzi}, updated 2011) for the lifetime
of the neutral pion is based
on six experiments: three relying on the Primakoff
effect~\cite{Kryshkin:1969wn,Bellettini:1970th,Browman:1974cu}, a direct
measurement~\cite{Atherton:1985av}, an $e^+e^-$ collider
measurement~\cite{Williams:1988sg}, and a measurement of the weak form
factor in $\pi^+\to e^+\nu\gamma$~\cite{Bychkov:2008ws}, which is
related to the $\pi^0$ lifetime via the conserved vector current hypothesis.
This leads to
the average lifetime $\tau_{\pi^0} = (8.4\pm 0.4)\times 10^{-17}s$.
Including the recent
precise measurement by PrimEx at JLab~\cite{Larin:2010kq},
\begin{equation}\label{Gpiggprimex}
 \Gamma(\pi^0\to\gamma\gamma)_\text{PrimEx} = 7.82\pm 0.14\pm0.17 \,\text{eV} \,,
\end{equation}
leads to a smaller uncertainty, $\tau_{\pi^0} = (8.35\pm 0.31)\times 10^{-17}s$.
Ongoing efforts by the PrimEx collaboration are expected 
to decrease the error by a factor two.

The partial decay width is related to the decay amplitude by
\begin{equation}
 \Gamma_{\gamma\gamma}  = \frac{M_\pi^3}{64 \pi} |F_{\pi\gamma\gamma}|^2\,.
\end{equation}
At lowest order we find from the Wess-Zumino-Witten Lagrangian
\begin{equation}
 F_{\pi\gamma\gamma}^\text{LO} = \frac{e^2}{4\pi^2 F_\pi}\quad \Rightarrow \quad
\Gamma(\pi^0\to\gamma\gamma)_\text{LO} \approx 7.76\,\text{eV}\,,
\end{equation}
which is in perfect agreement with the PDG average
as well as with the PrimEx result~\eqref{Gpiggprimex}.
Higher-order corrections might still
destroy the agreement and we can use
the leading logarithms to estimate the size of these contributions
and to examine the convergence of the chiral series. 
Notice that no chiral logarithms
are present at the one-loop level once everything is expressed in terms
of the physical quantities $F_\pi$
and $M_\pi$~\cite{Bijnens:1988kx,Donoghue:1986wv}. 
At the two-loop level, leading logarithms start to
contribute~\cite{Kampf:2009tk}.
At present the best prediction including electromagnetic and two-loop
effects is~\cite{Kampf:2009tk}
\begin{equation}
 \Gamma_{\pi^0\to\gamma\gamma} = (8.09 \pm 0.11)\,\text{eV}\,,
\end{equation}
which leads to the lifetime $\tau_{\pi^0} = (8.04 \pm 0.11)\,10^{-17}\,\text{s}$.

Our result for $\hat F$ in (\ref{Fgamma}) indicates that the convergence is fast
and higher orders are small. Putting in $\mu=0.77$~GeV we obtain
\begin{equation}
\hat F = 1+0-0.000372+0.000088+0.000036+0.000009+0.0000002+\ldots\;,
\end{equation}
which clearly shows a fast convergence.

We now turn to the discussion of the meson-photon transition
form factor $F_\gamma(-Q^2)$, normalized to the value at $Q^2=0$,
which has been given in (\ref{Fgamma}). 
It was measured by CELLO~\cite{Behrend:1990sr}, CLEO~\cite{Gronberg:1997fj},
and recently by BaBar~\cite{Aubert:2009mc}
mainly in the range $1 \leq Q^2 \leq 40~\text{GeV}^2$.
New activity is expected for very low $Q^2$ by KLOE-2 at DA$\Phi$NE~\cite{AmelinoCamelia:2010me} which
should directly test the prediction.

\begin{figure}[t]
\begin{center}
\includegraphics[angle=270,width=0.7\textwidth]{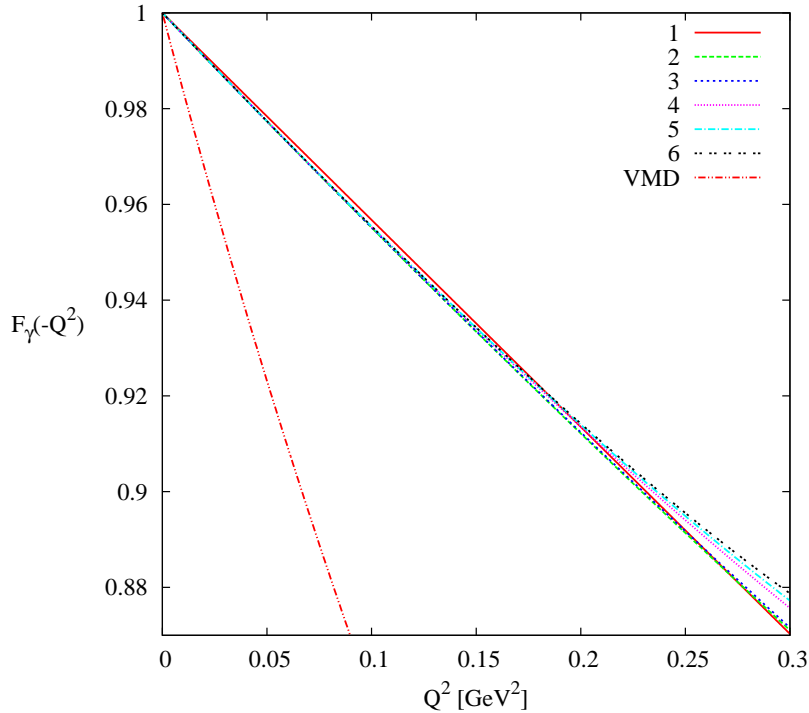}
\end{center}
\caption{The LL contribution to $F_\gamma(-Q^2)$ at different orders. Also shown is the
VMD prediction as a comparison.}
\label{figpiggofs}
\end{figure}

The LL contribution up to fifth order has been given in~(\ref{Fgamma}). 
Our result for the LL contribution to this
quantity is depicted in Fig.~\ref{figpiggofs}
together with the VMD prediction
\begin{equation}
F_{\gamma}^\text{VMD}(-Q^2) = \frac{m_V^2}{m_V^2+Q^2}\,. 
\end{equation}

\section{Conclusion} \label{conclusions}

In this paper we have extended the earlier work on leading logarithms in
effective field theories to the anomalous sector. First we improved the
programs used in the earlier work on the massive nonlinear sigma model
\cite{Bijnens:2009zi,Bijnens:2010xg}. This allowed us to compute one order
higher than in those papers and we presented results for the mass, decay
constant and the vector form factor. For the latter we clarified the
discrepancy with the chiral limit work of \cite{Kivel:2009az} and we
presented some numerical results as well.

The main part of this paper is the extension to the anomalous sector. We
thus added the Wess-Zumino-Witten term to the massive nonlinear sigma model for
$N=3$ and computed the leading logarithms to six-loop order
for $\pi^0\to\gamma^*\gamma^*$ and five-loop order for the $\gamma^*\pi\pi\pi$
vertex. We did not find a simple guess for the coefficients which was one of the
hopes when starting this work. In both cases the leading logarithms indicate
that the chiral series converges fast and we presented some numerical
results for the pion lifetime, the transition form factor and the
$\gamma^*\pi\pi\pi$ vertex.

\section*{Acknowledgments}
S.L.\ is supported by a grant from the Swiss National Science Foundation
and K.K.\ by the Center for Particle Physics (project No. LC 527) of the
Ministry of Education of the Czech Republic.
This work is supported in part by the European Community-Research
Infrastructure Integrating Activity ``Study of Strongly Interacting Matter'' 
(HadronPhysics2, Grant Agreement No. 227431)
and the Swedish Research Council grants 621-2008-4074 and 621-2010-3326.


\appendix

\section{Dispersive approach for the pion form factor}
\label{appdisform}

Since the leading logarithms for the vector form factor in the chiral
limit obtained in~\cite{Bijnens:2010xg} and here
do not agree with the corresponding result from
Kivel et al.~\cite{Kivel:2009az}, another check of our result is in order.
In~\cite{Koschinski:2010mr}, it was found that the partial wave amplitudes
for $\pi\pi$ scattering are given by
\begin{equation}\label{tIl}
 t^I_l(s)=\frac{\pi}{2}\sum_{n=1}^\infty \omega^{I}_{nl}\ \frac{\hat
S(s)^n}{2l+1}
\ln^{n-1}\Big(\frac{\mu^2}{s}\Big)
+\O(\text{NLL}),
\end{equation}
with the dimensionless function $\hat S(s) = \frac{s}{(4\pi F)^2}$. 
The coefficients $\omega^I_{nl}$ can be found in 
Tables~I and~II in~\cite{Koschinski:2010mr}.

The leading logarithms for the scalar form factor are
\begin{equation}
 F_S (s) = \sum_{n=0}^\infty f^S_n \hat S(s)^n
\ln^n\left(-\frac{\mu^2}{s}\right)\,.
\end{equation}
In this case, the results from~\cite{Kivel:2009az} and~\cite{Bijnens:2010xg} for
the coefficients $f^S_0$ are in agreement.
The discontinuity across the cut of the scalar form factor must satisfy
\begin{equation}
 \text{disc}\ F_S(s) = t^0_0 F_S(s) \,,
\end{equation}
which can be easily verified to hold for the coefficients $f^S_n$
given in~\cite{Kivel:2009az,Bijnens:2010xg}.

A similar expansion holds for the vector form factor:
\begin{equation}
 F_V (s) = \sum_{n=0}^\infty f^V_n \hat S(s)^n
\ln^n\Big(-\frac{\mu^2}{s}\Big) +\O(\text{NLL})\,.
\end{equation}
This time, however, the results from~\cite{Kivel:2009az}
disagree with ours and~\cite{Bijnens:2010xg}
for $n>2$.
The discontinuity across the cut of the vector form factor must hold
\begin{equation}
 \text{disc}\ F_V(s) = t^1_1 F_V(s)\,,
\end{equation}
which is only given for the $f_n^V$ from us and~\cite{Bijnens:2010xg}. We
therefore conclude that this is the correct result.

Dropping the factor $(-1)^{p+1}$ in (12) of~\cite{Kivel:2009az} brings
that result in agreement with ours. That there is indeed a
misprint in~\cite{Kivel:2009az}
was confirmed to us by the authors and was stated in
the PhD thesis of A.\ A.\ Vladimirov.

\begingroup\raggedright\endgroup

\end{document}